\documentstyle[12pt,psfig]{article}

\newcommand{\resection}[1]{\setcounter{equation}{0}\section{#1}}
\newcommand{\appsection}{\addtocounter{section}{1} \setcounter{equation}{0}
                         \section*{Appendix \Alph{section}}}

\newcommand{\EQ}{\begin{equation}}
\newcommand{\EN}{\end{equation}}
\newcommand{\bea}{\begin{eqnarray}}
\newcommand{\eea}{\end{eqnarray}}
\newcommand{\hs}{\hspace{0.1cm}}

\newcommand{\barr}{\overline}

\begin{document}
\setcounter{page}{0}
\topmargin 0pt
\oddsidemargin 5mm
\renewcommand{\thefootnote}{\arabic{footnote}}
\newpage
\setcounter{page}{0}
\begin{titlepage}
\begin{flushright}
ISAS/EP/98/63
\end{flushright}
\vspace{0.5cm}
\begin{center}
{\large {\bf Exact Matrix Elements in Supersymmetric
Theories}}\footnote{Work done under partial support of
the EC TMR Programme {\em Integrability, non--perturbative
effects
and symmetry in Quantum Field Theories}, grant FMRX-CT96-0012}
\\
\vspace{1.8cm} {\large G. Mussardo}\\ \vspace{0.5cm}
{\em International School for Advanced Studies,Trieste\\
and \\
Laboratoire de Physique Theorique et Hautes Energies, Paris VI}\\

\end{center}
\vspace{1.2cm}

\renewcommand{\thefootnote}{\arabic{footnote}}
\setcounter{footnote}{0}

\begin{abstract}
\noindent
The lowest representatives of the Form Factors relative to the trace
operators of $N=1$ Super Sinh-Gordon Model are exactly calculated. 
The novelty of their determination consists in solving a coupled
set of  unitarity and crossing equations. Analytic continuations of 
the Form Factors as functions of the coupling constant allows the 
study of interesting models in a uniform way, among these the latest 
model of the Roaming Series and the minimal supersymmetric models as 
investigated by Schoutens. A fermionic version of the $c$-theorem is also
proved and the  corresponding sum-rule derived.  
\end{abstract}

\vspace{1.5cm}



\end{titlepage}

\newpage
\noindent

\resection{Introduction}

The solution of a Quantum Field Theory (QFT) is provided by the complete
set of correlation functions of its fields. Perturbation theory often 
proves to be an inadequate approach to this problem therefore more
effective and powerful methods need to be developed. In this respect, 
one of the most promising methods is the Form Factor approach which is
applicable to integrable models \cite{KW,Smirnov}. This consists in
computing exactly all matrix elements of the quantum fields and then 
using them to obtain the spectral representations of the correlators. In
addition to the rich and interesting mathematical structure presented 
by the Form Factors themselves (which has been investigated in a series of
papers, among  which [2-13], 
the resulting spectral series usually show a remarkable 
convergent behaviour which allows approximation of the correlators (or 
quantities related to them) within any desired accuracy 
\cite{IMMF,YZ,ZamYL,CMpol,DM,CD,Saleur}.  

Particularly important examples of QFT are those which are invariant 
under a supersymmetry transformation which mixes the elementary bosonic 
and fermionic excitations. The subject of this paper is the investigation 
of the Form Factors of bidimensional supersymmetric theories, in
particular of the Super Sinh-Gordon Theory (SShG) and of those models 
which can be obtained by its analytic continuations. In these models, 
the degeneracy of the spectrum dictated by the supersymmetry implies 
the existence of multi--channel scattering processes and the resulting
$S$-matrix is necessarily non--diagonal. In this case, the
complete determination of the matrix elements of the quantum fields 
for an arbitrary number of asymptotic particles becomes a 
mathematical problem of formidable complexity. Although  
this problem has not been solved in its full generality in this paper, 
however some progress has been nevertheless achieved in determining the
lowest Form Factors of these non--diagonal scattering theories. 
In particular, a set of functions has been identified which might be used
in future studies to achieve the determination of all Form Factors of 
the models. As we will see, the calculation of the lowest matrix 
elements of supersymmetric theories presents some novelties which makes 
their determination an interesting issue in itself. In fact, for the 
first time one has to solve a coupled set of unitarity and crossing 
equations which originate from the fermionic nature of the supersymmetric 
theories. It should be said that in the past the calculation of the 
lowest Form Factors of supersymmetric theories has been approached 
in a paper by Ahn \cite{Ahn} but it turns out that his results were 
incorrect, as briefly discussed in Appendix B. 

The paper is organised as follows. The next section recalls    
the basic results of the scattering theories of bidimensional 
models with a $N=1$ supersymmetry. Section 3 deals with the 
simplest super-symmetric scattering theory which will be finally 
identified with a particular point of the ordinary Sine-Gordon 
$S$-matrix. The Super Sinh-Gordon Theory (SShG) and the relevant features 
of the deformed superconformal models are the subject of Section 4. 
The $S$-matrix of the SShG model and of others which 
can be obtained as analytic continuations thereof are analysed in 
Section 5. The determination of the lowest Form Factors of a 
particularly important class of operators -- those of the trace of 
the super stress-energy tensor -- is discussed in Section 6. In 
Section 7, a thorough check of their validity is obtained by 
means of the $c$-theorem sum rule as well as by comparing 
them with the limiting form obtained in two significant cases, the 
lastest representative of the Roaming Models and the first example of the
Schoutens's  models. Section 8 is devoted to the fermionic formulation 
of the $c$-theorem which can be achieved in supersymmetric 
theories whereas Section 9 contains the summary of the results and 
conclusions. There are two appendices, the first relates  
to the properties of a function entering the Form Factor calculation, 
the second contains a brief discussion and criticism of previous 
results obtained by Ahn. 
   
\resection{Generalities and Notation}

Scattering theories of integrable super-symmetric theories have 
been discussed in detail in Schoutens's paper  
\cite{Schoutens} (see also \cite{HM}). However, for the sake of 
clarity, in this section we explicitly state all notations and 
conventions which will be used in this paper. 

Let us consider a two-dimensional quantum field theory made of a
bosonic and fermion particle both of mass $m$. 
The one particle-state of the bosonic and fermionic particles will be 
denoted by $\mid b(\beta)\rangle$ and $\mid f(\beta)\rangle$ 
respectively\footnote{For the multi-particle states we have 
$\mid A_1(\beta_1) A_2(\beta_2) \cdots A_n(\beta_n)\rangle$, 
where each $A_i$ is either a $b$ or a $f$ particle and the rapidities 
are ordered in increasing order $\beta_1 > \beta_2 > \cdots \beta_n$ 
for the in-state and in a decreasing order $\beta_1 <\beta_2 < 
\cdots \beta_n$ for the out-state.}, where  
$\beta$ is the rapidity, i.e. the variable entering the dispersion
relations  $p_0 = m\cosh\beta$ and $p_1 = m\sinh\beta$ (in the following 
we will also consider the combinations $p_{\pm} =p_0 \pm p_1 =
m e^{\pm \beta}$). 

Let us assume that such a theory is both integrable (i.e. there exists an 
infinite number of conserved charges) and invariant under a $N=1$ 
supersymmetry. This means that among the set of conserved quantities 
there are two fermionic charges ${\cal Q}$ and $\barr{\cal Q}$ which 
satisfy the relations 
\EQ
{\cal Q}^2 = P_{+} 
\,\,\, ,\,\,\, 
\barr {\cal Q}^2 = P_{-}  
\,\,\, ,\,\,\, 
\{{\cal Q},\barr{\cal Q}\} =0 
\,\,\,, 
\label{algebra}
\EN 
where $P_{\pm}$ are the right/left components $P_{\pm}=P_0 \pm P_1$ 
of the momentum operator. The action of these charges on the one-particle 
states can be represented as 
\EQ
\begin{array}{l}
{\cal Q} \mid b(\beta) \rangle = \omega \,\,\,\,\sqrt{m} e^{\beta/2} 
\mid f(\beta)\rangle \,\,\,;\\ 
{\cal Q} \mid f(\beta) \rangle = \omega^{-1} \sqrt{m} e^{\beta/2} 
\mid b(\beta)\rangle\,\,\,;
\end{array}
\,\,\, \,\,\, 
\begin{array}{l}
\barr {\cal Q} \mid b(\beta) \rangle = \delta \,\,\,\,\sqrt{m} e^{-\beta/2} 
\mid f(\beta)\rangle \,\,\,;\\ 
\barr {\cal Q} \mid f(\beta) \rangle = \delta^{-1} \sqrt{m} e^{-\beta/2} 
\mid b(\beta)\rangle \,\,\,,
\end{array}
\label{Qoneparticle}
\EN 
i.e. in terms of two matrices 
\EQ
Q = \left(
\begin{array}{cc} 
0 & \omega \\
\omega^{-1} & 0 \\
\end{array}
\right)
\,\,\,\,\, , \,\,\,\,\, 
\barr Q = \left(
\begin{array}{cc} 
0 & \delta \\
\delta^{-1} & 0 \\
\end{array}
\right) 
\,\,\, ,
\EN 
satisying $Q^2 = \barr Q^2 =1$. The anti-commutativity of the operators 
${\cal Q}$ and $\barr {\cal Q}$ gives the condition 
\EQ 
\omega = \pm i \delta \,\,\, ,
\label{alphadelta}
\EN 
with the actual values of $\omega$ and $\delta$ which will be fixed by 
later considerations. 

The action of ${\cal Q}$ and $\barr {\cal Q}$ on a multi-particle 
states must take into account the fermionic nature of these operators 
and therefore involves brading relations. In the notation of 
ref.\cite{HM}, we have 
\begin{eqnarray}
&& {\cal Q}  
\mid A_1(\beta_1) A_2(\beta_2) \ldots A_n(\beta_n) \rangle = 
\sqrt{m} \sum_{k=1}^n e^{\beta_k/2} 
\label{braidingQ} 
\\ 
&& \,\,\,
\mid (Q_F A_1(\beta_1)) (Q_F A_2(\beta_2)) \ldots 
(Q_F A_{k-1}(\beta_{k-1}) (Q A_k(\beta_k))
A_{k+1}(\beta_{k+1}) \ldots A_n(\beta_n) \rangle  
\nonumber
\end{eqnarray}
and 
\begin{eqnarray}
&& \barr {\cal Q} 
\mid A_1(\beta_1) A_2(\beta_2) \ldots A_n(\beta_n) \rangle = 
\sqrt{m} \sum_{k=1}^n e^{-\beta_k/2} 
\label{braidingQbar} \\ 
&& \,\,\,
\mid (Q_F A_1(\beta_1)) (Q_F A_2(\beta_2)) \ldots 
(Q_F A_{k-1}(\beta_{k-1}) (\barr Q A_k(\beta_k))
A_{k+1}(\beta_{k+1}) \ldots A_n(\beta_n) \rangle  
\nonumber
\end{eqnarray}
where $Q_F$ is the fermion parity operator, which on the basis 
$\mid b\rangle$ and $\mid f\rangle$ is represented by the 
diagonal matrix 
\EQ
Q_F = \left(\begin{array}{cc}
1 & 0 \\
0 & -1 
\end{array}
\right) \,\,\, . 
\label{Q_L}
\EN  
Particularly important is the representation of the two super-charges  
on the two-particle states $\mid b(\beta_1) b(\beta_2)\rangle$,  
$\mid f(\beta_1) f(\beta_2)\rangle$, $\mid f(\beta_1) b(\beta_2)\rangle$, 
$\mid b(\beta_1) f(\beta_2)\rangle$. The first two states belong to the 
$F=1$ sector whereas the remaining two states to the $F=-1$ sector. 
By choosing for convenience $\beta_1 =\beta/2$ and $\beta_2 =-\beta/2$, 
the operator ${\cal Q}$ will be represented by the matrix 
\EQ
Q(\beta) = 
\left(\begin{array}{cccc}
0 & 0 & \omega x & \omega x^{-1} \\
0 & 0 & -\omega^{-1} x^{-1} & \omega^{-1} x \\
\omega^{-1} x & -\omega x^{-1} & 0 & 0 \\
\omega^{-1} x^{-1} & \omega x & 0 & 0 
\end{array}
\right) 
\,\,\, ,
\label{q2particle}
\EN 
where $x\equiv e^{\beta/4}$. For $\barr{\cal Q}$ we have analogously 
\EQ
\barr Q(\beta) = 
\left(\begin{array}{cccc}
0 & 0 & \delta x^{-1} & \delta x \\
0 & 0 & -\delta^{-1} x & \delta^{-1} x^{-1} \\
\delta^{-1} x^{-1} & -\delta x & 0 & 0 \\
\delta^{-1} x & \delta x^{-1} & 0 & 0 
\end{array}
\right) \,\,\,.
\label{qbar2particle}
\EN 
In the following we will also need the representation 
matrix of the operator ${\cal Q}\barr{\cal Q}$ on the above 
two particle states, given by  
\EQ
(Q \barr Q)(\beta) = 2  
\left(\begin{array}{cccc}
\frac{\omega}{\delta} & -\omega \delta \sinh\frac{\beta}{2}& 0 & 0 \\
\frac{1}{\omega\delta} \sinh\frac{\beta}{2} & -\frac{\omega}{\delta} & 0 & 0\\
0 & 0 & 0 & -\frac{\omega}{\delta} \cosh\frac{\beta}{2} \\
0 & 0 & -\frac{\omega}{\delta} \cosh\frac{\beta}{2} & 0 
\end{array}
\right) \,\,\,.
\label{qqbarparticle}
\EN 

Let us consider now the structure of the elastic two-body $S$-matrix 
of the theory. Since fermions can only created or destroyed in couples, 
we have the following scattering channels 
\begin{eqnarray}
\mid b(\beta_1) b(\beta_2) \rangle & = &  
A(\beta_{12}) \mid b(\beta_2) b(\beta_1)\rangle + 
B(\beta_{12}) \mid f(\beta_2) f(\beta_1) \rangle \,\,\,;\nonumber \\
\mid f(\beta_1) f(\beta_2) \rangle & = &  
C(\beta_{12}) \mid b(\beta_2) b(\beta_1)\rangle + 
D(\beta_{12}) \mid f(\beta_2) f(\beta_1)\rangle 
\label{scatteringmatrix} \,\,\,;\\
\mid f(\beta_1) b(\beta_2) \rangle & = &
E(\beta_{12}) \mid f(\beta_2) b(\beta_1)\rangle + 
F(\beta_{12}) \mid b(\beta_2) f(\beta_1) \rangle 
\nonumber \,\,\,;\\
\mid b(\beta_1) f(\beta_2) \rangle & = &
G(\beta_{12}) \mid f(\beta_2) b(\beta_1) \rangle + 
H(\beta_{12}) \mid b(\beta_2) f(\beta_1)\rangle \,\,\, ,\nonumber 
\end{eqnarray} 
where $\beta_{12} = \beta_1 -\beta_2$. Respecting the ordering of the
2-particle states as above, the $S$-matrix can be then represented by 
\EQ 
S(\beta) = \left(\begin{array}{cccc} 
A & B & 0 & 0 \\
C & D & 0 & 0 \\
0 & 0 & E & F \\
0 & 0 & G & H 
\end{array}
\right) \,\,\,. 
\label{matrixs}
\EN 
The amplitudes $A(\beta)$ and $D(\beta)$ describe the transmission 
channels of two bosons and two fermions respectively whereas 
the amplitudes $B(\beta)$ and $C(\beta)$ are related to 
the annihilation-creation channels of two bosons into two fermions 
and viceversa. The other amplitudes $E(\beta)$ and $H(\beta)$ 
describe the reflection channels of the boson-fermion scattering 
whereas the remaining $F(\beta)$ and $G(\beta)$ describe 
the pure transmission processes in the $F=-1$ sector. 

Let us now impose the invariance of the scattering processes under 
the action of the supersymmetric charges. First consider the 
invariance of the $S$-matrix with respect to the operator 
${\cal Q}\barr{\cal Q}$: with the choice $\beta_1 = \beta/2$ and 
$\beta_2 = -\beta/2$, by taking into account the different 
ordering of the rapidities of the out-states, this implies
the following constraint on the scattering amplitudes 
\cite{Schoutens}
\EQ
Q \barr Q(\beta) S(\beta) = S(\beta) (Q \barr Q)(-\beta)
\,\,\, .
\label{1invariance}
\EN 
In the $F=-1$ sector, this equation leads to the conditions  
\EQ
E(\beta) = H(\beta) \,\,\,\,\, , \,\,\,\,\, 
F(\beta) = G(\beta) \,\,\, , 
\label{1constraint}
\EN 
whereas in the $F=1$ sector, eq.(\ref{1invariance}) implies 
\begin{eqnarray}
B(\beta) & = & (\omega \delta)^2 \, C(\beta) \,\,\,;\nonumber \\
2 \omega^2 \,C(\beta) & = & [A(\beta) + D(\beta)] \sinh\frac{\beta}{2} 
\label{2constraint} \,\,\, ;\\
\frac{2}{\delta^2}\,B(\beta) & = & [A(\beta) + D(\beta)] \sinh\frac{\beta}{2}
\nonumber \,\,\,.
\end{eqnarray}
Hence, we can choose $B(\beta)=C(\beta)$ provided that the constants
$\omega$ and $\delta$ satisfy the condition   
\EQ
(\omega\delta)^2 =1 \,\,\,.
\label{alphadeltacond}
\EN 
Since from the closure of the supersymmetric algebra 
$\omega = \pm i \delta$, it follows that $\delta$ has to satisfy 
the condition $\delta^4 =-1$ and therefore we have the following 
possibilities: 
\EQ
\delta = 
e^{i \frac{\pi}{4}} 
\,\,\,\,\,\,\, ,
\,\,\,\,\, 
\omega = \pm e^{-i \frac{\pi}{4}} \,\,\,.
\label{pi/41}
\EN 
or 
\EQ
\delta = 
e^{-i \frac{\pi}{4}} 
\,\,\,\,\,\,\, ,
\,\,\,\,\, 
\omega= \pm e^{i \frac{\pi}{4}} \,\,\,.
\label{pi/42}
\EN 
In the following we will adopt the values $\omega=e^{i \frac{\pi}{4}}$ 
and $\delta = e^{-i \frac{\pi}{4}}$. With this choice, 
we have then 
\EQ
B(\beta) = C(\beta) 
\,\,\,\,\, ,
\,\,\,\,\, 
2 B(\beta) = - i\,[A(\beta) + D(\beta)]\, \sinh\frac{\beta}{2} \,\,\,.
\label{finalcon}
\EN 
In light of all the above equations, the $S$ matrix can be written 
then as 
\EQ 
S(\beta) = \left(\begin{array}{cccc} 
A & B & 0 & 0 \\
B & D & 0 & 0 \\
0 & 0 & E & F \\
0 & 0 & F & E 
\end{array}
\right) \,\,\,.
\label{matrixsfin}
\EN 
Further conditions on the above functions are provided 
by the invariance of the $S$-matrix under each supercharge separately
(having already considered the invariance under ${\cal Q}\barr{\cal Q}$, 
one can consider only one of them, say ${\cal Q}$) in the form 
\EQ
Q(\beta) S(\beta) = S(\beta) Q(-\beta) \,\,\,.
\label{2invariance}
\EN
This equation is equivalent to the following conditions 
\begin{eqnarray}
A(\beta) - D(\beta) & = & 2 F(\beta) \,\,\,;\nonumber \\
A(\beta) + D(\beta) & = & \frac{2}{\cosh\frac{\beta}{2}}\, E(\beta)
\label{mixingcond} \,\,\,;\\
E(\beta) \tanh\frac{\beta}{2} & = & i\,B(\beta) \,\,\,.\nonumber 
\end{eqnarray}
Finally, crossing symmetry implies 
\begin{eqnarray}
&& A(i \pi -\beta) = A(\beta) \nonumber \,\,\,;\\
&& D(i \pi -\beta) = D(\beta) \label{crossing} \,\,\,;\\
&& F(i \pi -\beta) = F(\beta) \nonumber \,\,\,;\\
&& E(i \pi -\beta) = B(\beta)\nonumber \,\,\,, 
\end{eqnarray}
whereas the unitarity condition $S(\beta) S(-\beta)=1$ gives  
\begin{eqnarray}
&& A(\beta) A(-\beta) + B(\beta) B(-\beta) = 1 \,\,\,;\nonumber \\
&& A(\beta) B(-\beta) + B(\beta) D(-\beta) = 0 \,\,\,;\nonumber \\
&& B(\beta) A(-\beta) + D(\beta) B(-\beta) = 0 \,\,\,;\nonumber \\
&& B(\beta) B(-\beta) + D(\beta) D(-\beta) = 1 \,\,\,;\label{unitarity} \\
&& E(\beta) E(-\beta) + F(\beta) F(-\beta) = 1 \,\,\,;\nonumber \\
&& E(\beta) F(-\beta) + F(\beta) E(-\beta) = 0 \,\,\,.\nonumber 
\end{eqnarray}

In the next section we will discuss several examples of scattering 
theories which fulfill the above set of equations. 

\resection{The simplest SUSY S-matrix} 

The simplest supersymmetric $S$-matrix can be obtained by noticing 
that the amplitude $F(\beta)$ is invariant under crossing. This 
allows us to make the consistent choice $F(\beta)=0$. In view of the 
equations (\ref{mixingcond}), the $S$ matrix can be written in 
this case as 
\EQ 
S(\beta) = \left(\begin{array}{ccrr} 
-1/\cosh\frac{\beta}{2} 
& i \tanh\frac{\beta}{2} 
& 0 & 0 \\
i \tanh\frac{\beta}{2} & 
-1/\cosh\frac{\beta}{2} &
0 & 0 \\
0 & 0 & -1 & 0 \\
0 & 0 & 0 & -1 
\end{array}
\right) \,R(\beta) \,\,\,, 
\label{matrixssim}
\EN 
where the function $R(\beta)$ has to satisfy the equations 
\EQ
\begin{array}{l}
R(\beta) R(-\beta) =1 \,\,\, ;\\
R(i \pi -\beta) = -i \tanh\frac{\beta}{2} \, R(\beta) \,\,\,. 
\end{array}
\label{unicrosmin}
\EN 
There is a geometrical method of solving the above coupled set of 
equations for a meromorphic function with only poles and zeros, none 
of which are in the physical strip $0 \leq {\rm Im}\,\beta \leq \pi$. 
First of all, the poles and zeros of $R(\beta)$ are linked to each 
other by the first eq. of (\ref{unicrosmin}) because if the function 
$R(\beta)$ has a pole (zero) in $i \pi \eta$, it should necessary have 
a zero (pole) in $-i \pi\eta$. On the other hand, changing the 
rapidity $\beta$ according to $\beta \rightarrow i \pi -\beta$ is 
a reflection with respect to the point $i \pi$. 
The function $-i\tanh\frac{\beta}{2}$ has zeros in $\beta =\pm 2 k \pi i$ 
and poles in $\beta=\pm (k+1)\pi i$ ($k=0,1,\ldots$) and its infinite 
product representation is given by 
\EQ
-i\tanh\frac{\beta}{2} = \prod_{k=0}^{\infty} 
\frac{[2 k \pi -i\beta][(2 (k+1) \pi + i \beta]}
{[(2 k+1)\pi +i\beta][(2 k+1) \pi -i \beta]}
\label{productth} \,\,\,.
\EN 
Since the second equation can be written as 
\EQ
R(i \pi -\beta) R(-\beta) = -i \tanh\frac{\beta}{2} \,\,\, ,  
\EN
to solve this equation we therefore need to find a function 
whose sovraposition of poles and zeros coming from the combination 
of the two reflections $\beta \rightarrow i\pi-\beta$ 
and $\beta \rightarrow -\beta$ match those of (\ref{productth}). 
It is simple to see that such a function should have zeros and poles 
of increasing multiplicities and its infinite product representation 
is given by 
\EQ
R(\beta) = \prod_{k=0}^{\infty} 
\left( 
\frac
{[(2 k+1) \pi + i \beta][(2 k+3) \pi + i \beta][2 (k+1)\pi -i\beta]^2} 
{[(2 k+1) \pi - i \beta][(2 k+3) \pi - i \beta][2 (k+1)\pi +i\beta]^2} 
\right)^{k+1} \,\,\, .
\label{1represeR}
\EN 
The above function can also be written as 
\EQ
R(\beta) = \prod_{k=0}^{\infty} 
\frac{
\Gamma\left(k+\frac{1}{2} -i\frac{\beta}{2\pi}\right)
\Gamma\left(k+\frac{3}{2} -i\frac{\beta}{2\pi}\right)
\Gamma^2\left(k+1 +i\frac{\beta}{2\pi}\right)}
{\Gamma\left(k+\frac{1}{2} +i\frac{\beta}{2\pi}\right)
\Gamma\left(k+\frac{3}{2} +i\frac{\beta}{2\pi}\right)
\Gamma^2\left(k+1 -i\frac{\beta}{2\pi}\right)} \,\,\, ,
\label{repreRGamma}
\EN 
or as 
\EQ
R(\beta) = \exp\left[\frac{i}{2} \int_0^{\infty} \frac{dt}{t} 
\frac{\sin\frac{\beta t}{\pi}}{\cosh^2\frac{t}{2}}
\right]  \,\,\,.
\label{repreRint}
\EN 
It is now interesting to note that the simplest supersymmetric 
$S$-matrix as above coincides with the $S$-matrix coming 
from another integrable model. Consider, in fact, the $S$-matrix 
in the solitonic sector of the ordinary Sine-Gordon model, given in 
full generality by\footnote{The two-particle states are ordered 
as $\mid S\barr S\rangle$, $\mid \barr S S\rangle$, $\mid S S\rangle$, 
$\mid \barr S\barr S\rangle$, where $S$ denotes the soliton whereas 
$\barr S$ the anti-soliton.} \cite{ZZ}  
\EQ 
S(\beta) = \hat R(\beta)\,\left(\begin{array}{cccc} 
\frac{i\sin\frac{\pi^2}{\xi}}{\sinh\frac{\pi (\beta-i\pi)}{\xi}} &
\frac{\sinh\frac{\pi \beta}{\xi}}{\sinh\frac{\pi(\beta-i\pi)}{\xi}} &
0 & 0 \\ 
\frac{\sinh\frac{\pi \beta}{\xi}}
{\sinh\frac{\pi (\beta-i\pi}{\xi}}
 & \frac{i\sin\frac{\pi^2}{\xi}}{\sinh\frac{\pi (\beta-i\pi)}{\xi}} & 
0 & 0 \\0 & 0 & -1 & 0 \\
0 & 0 & 0 & -1 
\end{array}
\right) \,\,\, ,
\label{sgsmatrix}
\EN 
where 
\EQ
\hat R(\beta) = \exp\left[-i\int_0^{\infty}\frac{dt}{t} 
\frac{\sinh\frac{t}{2}\left(1-\frac{\xi}{\pi}\right)}
{\cosh\frac{t}{2} \sinh\frac{\xi t}{2 \pi}} 
\sin\frac{\beta t}{\pi}\right] \,\,\,.
\label{USG}
\EN 
The parameter $\xi$ is related to the coupling constant $g$ 
of the Lagrangian of the model  
\EQ
{\cal L} =\frac{1}{2} (\partial_{\mu} \varphi)^2 +  
\frac{m^2}{g^2} \cos(g \varphi) \,\,\, ,
\label{SG}
\EN 
by $\xi=\frac{\pi g^2}{8 \pi -g^2}$. 

By comparing the $S$-matrix (\ref{matrixssim}) with the one 
of the Sine-Gordon model, it is now easy to see that the two  
coincide for the particular value $\xi = 2\pi$, i.e. $g^2=16\pi/3$. 
This is a repulsive point of the Sine-Gordon model, no additional 
bound states are present at this value and therefore the full $S$-matrix 
of the SG model is reduced to the one in (\ref{sgsmatrix}). Since the
ordinary Sine-Gordon model can be regarded as a massive integrable 
deformation of the gaussian conformal model with an action 
\EQ
{\cal A}_0 = \frac{1}{2} \int d^2x (\partial_{\mu}\varphi)^2 \,\,\, ,
\label{gaussian}
\EN
(of central charge $C=1$), the deforming operator 
$(e^{i g\varphi} + e^{-i g\varphi})$ which leads to the Sine-Gordon model 
has at the point $\xi = 2\pi$ the conformal dimension $\Delta=2/3$. 
The relevance of these observations as well as the consequences of the 
identity between the $S$-matrices will be topics of discussion
later in this paper. 

\resection{The SShG and Superconformal Models}

The aim of this section is to illustrate several properties of the 
Super Sinh-Gordon model as well as some features of the superconformal 
models and their deformations. 

\subsection{Lagrangian of the SShG model}

In the euclidean space, the Super Sinh-Gordon model can be
defined in terms of its action 
\EQ
{\cal A} = \int d^2z d^2\theta \left[
\frac{1}{2} D\Phi \barr D \Phi + i \frac{m}{\lambda^2} 
\cosh\lambda\Phi \right] \,\,\, ,
\label{actionSSHG}
\EN 
where the covariant derivatives are defined as 
\EQ
\begin{array}{l}
D = \partial_{\theta} - \theta \,\partial_z \,\,\,;\\
\barr D = \partial_{\barr\theta} -\barr\theta \,\partial_{\bar z}\,\,\, ,
\end{array}
\label{covariant}
\EN 
and the superfield $\Phi(z,\bar z,\theta,\barr\theta)$ has an expansion as 
\EQ
\Phi(z,\bar z,\theta,\barr\theta)= \varphi(z,\bar z) + 
\theta \psi(z,\bar z) + \barr \theta \,\barr \psi(z,\bar z) + 
\theta \barr \theta \,{\cal F}(z,\bar z) \,\,\,.
\label{superfield}
\EN 
The integration on the $\theta$ variables as well as the elimination of 
the auxiliary field ${\cal F}(z,\bar z)$ by means of its algebraic equation 
of motion leads to the Lagrangian 
\EQ
{\cal L} = \frac{1}{2} 
(\partial_z\varphi \partial_{\bar z}\varphi + 
\barr\psi \partial_z \barr\psi + \psi\partial_{\bar z} \psi ) 
+\frac{m^2}{2 \lambda^2} \sinh^2\lambda\varphi + 
i m \barr\psi \psi \, \cosh\lambda\varphi \,\,\, .
\label{lagrangianSSHG}
\EN 

Of all the different ways of looking at the SSHG model, one of 
the most convenient is to consider it as a deformation of the 
superconformal model described by the action 
\EQ
{\cal A_0} = \frac{1}{2}\int(  
\partial_z\varphi \partial_{\bar z}\varphi + 
\barr\psi \partial_z \barr\psi + \psi\partial_{\bar z} \psi) \,\,\,.
\label{SCFT3/2}
\EN
This superconformal model has central charge $C=3/2$. At this poit, 
it is useful to briefly remind some properties of the superconformal 
models and their deformations.  

\subsection{Superconformal models and their deformation}

For a generic superconformal model, the supersymmetric 
charges can be represented by the differential operators 
\EQ
\begin{array}{l}
Q = \partial_{\theta} + \theta \,\partial_z \,\,\,;\\
\overline Q = \partial_{\bar\theta} +\barr\theta \,\partial_{\bar z} \,\,\,.
\end{array}
\label{differentialcharges}
\EN 
The analytic part of the stress-energy tensor $T(z)$ and the 
current $G(z)$ which generates the supersymmetry combine 
themselves into the analytic superfield 
\EQ
W(z,\theta) = G(z) + \theta T(z) \,\,\, ,
\label{SSET}
\EN
which is called the super stress-energy tensor. 
For the anti-analytic sector we have correspondingly 
$\barr W(\bar z) = \barr G(\bar z) +\barr\theta \,\barr T(\bar z)$. 
These fields are mapped one into the other by means of the 
super-charges 
\EQ
\begin{array}{lll}
T(z) = \{G(z),Q\} &,& \partial_z G =[T(z),Q] \,\,\,;\\
\barr T(\bar z) = \{\barr G(\bar z),\barr Q\} &,& 
\partial_{\bar z} \barr G =[\barr T(\bar z),\barr Q] \,\,\, , 
\end{array}
\label{susytransf}
\EN
and their Operator Product Expansion reads 
\begin{eqnarray}
T(z) T(w) &=& \frac{C}{2(z-w)^4} + 2 \frac{T(w)}{(z-w)^2} + 
\frac{\partial T(w)}{z-w} + \cdots \nonumber \\
T(z) G(w) &=& \frac{3}{2} \frac{G(w)}{(z-w)^2} +  
\frac{\partial G(w)}{z-w} + \cdots \label{OPE}\\
G(z) G(w) &=& \frac{2C}{3(z-w)^3} + 2 \frac{T(w)}{(z-w)} + 
\cdots \nonumber 
\end{eqnarray}
with analogous relations for the anti-analytic fields. 
As it is well known \cite{superconformal}, reducible unitary
representations of the $N=1$ superconformal symmetry occurs 
for the discrete values of the central charge 
\EQ
C = \frac{3}{2} -\frac{12}{m (m+2)} \,\,\,. 
\label{discretec}
\EN 
At these values, realizations of the $N=1$ superconformal algebra 
are given in terms of a finite number of superfields in the Neveu-Scwartz 
sector and a finite number of ordinary conformal primary fields 
in the Ramond sector. Their conformal dimensions are given by 
\EQ
\Delta_{p,q} = \frac{[(m+2) p -m q]^2-4}{8 m (m+2)} +
\frac{1}{32} [1 -(-1)^{p-q}] \,\,\, ,
\label{anomalousdimensions}
\EN 
where $p-q$ even corresponds to the primary Neveu-Schwartz 
superfields ${\cal N}_{p,q}^{(m)}(z,\theta)$ and $p-q$ odd to the 
primary Ramond fields $R_{p,q}^{(m)}(z)$. These fields enter the 
so-called superconformal minimal models ${\cal SM}_m$.  

The Witten index $Tr (-1)^F$ of the superconformal models can be 
computed by initially defining them on a cylinder \cite{KMS}: the 
Hamiltonian on the cylinder is given by $H=Q^2=L_0-C/24$, where 
as usual $C/24$ is the Casimir energy on the cylinder and 
$L_0 = \frac{1}{2\pi i}\oint dz z T(z)$. Considering that for any
conformal state $\mid a\rangle$ with $\Delta > C/24$ there is the 
companion state $Q\mid a\rangle$ of opposite fermionic parity, their 
contributions cancel each other in $Tr (-1)^F$ and therefore 
only the ground states with $\Delta = C/24$ (which are not 
necessarily paired) enter the final expression of the Witten 
index. For the minimal models, there is a non-zero 
Witten index only for $m$ even. Therefore the lowest superconformal 
minimal model with a non-zero Witten index is the one with 
$m=4$, which has a central charge $C=1$ and corresponds to 
the class of universality of the critical Ashkin-Teller model. 
The superconformal theory with $C=3/2$ made of free 
bosonic and fermionic fields also has a non-zero Witten index, 
because an unpaired Ramond field $R(z)$ is explicitly given by the spin 
field $\sigma(z)$ of Majorana fermion $\psi(z)$ with conformal 
dimension $\Delta =1/16$, i.e. by the magnetization operator of 
the Ising model.  

The above observations become important in the understanding 
the off-critical dynamics relative to the deformation of 
the action of the superconformal minimal models ${\cal SM}_m$ by 
means of the relevant supersymmetric Neveu-Schwartz operator 
${\cal N}_{1,3}^{(m)}(z,\theta)$. In fact, as shown in \cite{KMS}, 
the massless Renormalization Group flow generated by such an operator 
preserves the Witten index. Therefore the long distance 
behaviour of the deformed ${\cal SM}_m$ minimal model -- controlled 
by the action 
\EQ
{\cal A}_m + \gamma \int d^2z \,d^2\theta \,{\cal N}_{1,3}^{(m)} \,\,\, ,
\label{deformationm}
\EN 
(for a positive value of the coupling constant $\gamma$, with 
the usual conformal normalization of the superfield) -- is ruled by 
the fixed point of the minimal model ${\cal SM}_{m-2}$. Therefore 
the action (\ref{deformationm}) describes the RG flow 
\EQ
{\cal SM}_{m} \rightarrow {\cal SM}_{m-2} \,\,\, ,
\label{jump2}
\EN 
with a corresponding jump in steps of two of the central 
charge\footnote{This is in contrast of what happens in 
the ordinary conformal minimal models, where the deformation 
of the conformal action by means of the operator $\phi_{1,3}$ induces 
a massless flow between two next neighborod minimal models.}, i.e. 
$$\Delta C =C(m)-C(m-2) \,\,\, .$$ 
Therefore, a cascade of massless flows which start  
from $C=3/2$ and progress by all ${\cal N}_{1,3}^{(m)}$
deformations of ${\cal SM}_m$ met along the way must  
necessarily pass through the model with $C=1$ in the second to last 
step rather than the lowest model\footnote{As well known, the lowest 
model with $C=7/10$ corresponds to the class of universality of the 
Tricritical Ising Model.} with $C=7/10$ (Figure 1). We will see that 
this is indeed the scenario which is described by a specific 
analytical continuation of the coupling constant of the 
Super-Sinh-Gordon model, the so-called Roaming Models. 

For the time being, let us further discuss the deformation of 
the $C=3/2$ superconformal theory which leads to 
the SShG model. At the conformal point, the explicit realization 
of the component of the super stress-energy tensor are given by 
\EQ
\begin{array}{l}
T(z) = -\frac{1}{2}\left[
(\partial_z\varphi)^2 - \psi \partial\psi\right] \,\,\, ;\\
G(z) = \,i\psi \partial_z \varphi \,\,\, ,
\end{array}
\label{superstressc3/2}
\EN 
and they satisfy the conservation laws 
$\partial_{\bar z} T(z) = \partial_{\bar z} G(z) =0$. 
Once this superconformal model is deformed according to the Lagrangian 
(\ref{lagrangianSSHG}), the new conservation laws are given by 
\EQ
\begin{array}{l}
\partial_{\bar z} T(z,\bar z) =\partial_z \Theta(z,\bar z)
\,\,\, ;\\
\partial_{\bar z} G(z,\bar z) = \partial_z \chi(z,\bar z)  \,\,\, ,
\end{array}
\label{conservationlawsuperstress}
\EN 
where 
\begin{eqnarray}
&& \Theta(z,\bar z) = \frac{m^2}{2 \lambda^2} \sinh^2\lambda\varphi + 
i m \barr\psi \psi \, \cosh\lambda\varphi \,\,\, ,
\label{tracethetag}\\
&& \chi(z,\bar z) = \,\frac{m}{\lambda}\barr \psi \,\sinh\lambda\varphi
\,\,\, .\nonumber 
\end{eqnarray}
For the anti-analytic part of the super stress-energy tensor
we have 
\EQ
\begin{array}{l}
\partial_{z} \bar T(z,\bar z) =\partial_{\bar z} \Theta(z,\bar z)
\,\,\, ;\\
\partial_{z} \barr G(z,\bar z) = \partial_{\bar z}\barr \chi(z,\bar z) 
\,\,\, , \end{array}
\label{conservationlawantisuperstress}
\EN 
where $\Theta(z,\bar z)$ is as before and the other fields are given 
by 
\begin{eqnarray}
&& \barr G(z,\bar z) = -i \barr \psi \partial_{\bar z} \varphi \,\,\,;
\label{tracethetag2} \\
&& \barr\chi(z,\bar z) = \,\,\frac{m}{\lambda}\psi \,\sinh\lambda\varphi
\,\,\, .\nonumber 
\end{eqnarray}
The operators $\Theta(z,\bar z)$, $\chi(z,\bar z)$ and 
$\barr\chi(z,\bar z)$ belong to the trace of the supersymmetric 
stress-energy tensor and they are related each other by 
\EQ
\begin{array}{lll}
\Theta(z,\bar z) = \{\chi(z,\bar z),{\cal Q}\} &,& 
\partial_z \chi(z,\bar z) = [\Theta(z,\bar z),{\cal Q}]\,\,\,; \\
\Theta(z,\bar z) = \{\barr \chi(z,\bar z),\barr {\cal Q}\} &,& 
\partial_{\bar z} \barr \chi(z,\bar z) =[\Theta(z,\bar z),\barr 
{\cal Q}] \,\,\, ,
\end{array}
\label{susytransftrace}
\EN
where the charges of supersymmetry are expressed by 
\EQ
\begin{array}{l}
Q = \int G(z,\bar z) dz + \chi(z,\bar z) d\bar z\,\,\, ;\\
\barr Q = \int \barr G(z,\bar z) d\bar z + \bar \chi(z,\bar z) dz 
\,\,\, .
\end{array}
\label{integralcharges}
\EN 
In addition to the conservation laws (\ref{conservationlawsuperstress}) 
and (\ref{conservationlawantisuperstress}), the SShG model possesses 
higher integrals of motion which were explicitly determined in 
\cite{FGS}. Therefore its scattering processes are purely elastic 
and factorizable. and its two-body $S$-matrix is discussed in the 
next section. 

\resection{The $S$-matrix of the SSHG}

The $S$-matrix of the SSHG model has been determined in 
\cite{Ahn}. It is given by 
\EQ 
S(\beta) = Y(\beta)\,
\left(\begin{array}{cccc} 
1-\frac{2 i \sin\pi\alpha}{\sinh\beta} & 
\frac{-\sin\pi\alpha}{\cosh\frac{\beta}{2}} 
& 0 & 0 \\ 
\frac{-\sin\pi\alpha}{\cosh\frac{\beta}{2}} & 
-1 -\frac{2 i \sin\pi\alpha}{\sinh\beta} & 0 & 0 \\
0 & 0 &-\frac{i\sin\pi\alpha}{\sinh\frac{\beta}{2}} & 1 \\
0 & 0 & 1 & 
-\frac{i\sin\pi\alpha}{\sinh\frac{\beta}{2}}
\end{array}
\right) 
\label{SmatSSHG}\\
\EN 
where 
\EQ
Y(\beta) = \frac{\sinh\frac{\beta}{2}}{\sinh\frac{\beta}{2} + 
i \sin\pi\alpha} \, U(\beta,\alpha) \,\,\, ,
\EN 
and the function $U(\beta)$ is given by 
\EQ
U(\beta) = \exp\left[i 
\int_0^{\infty} \frac{dt}{t} 
\frac{\sinh\alpha t\,\sinh(1-\alpha)t}{\cosh^2\frac{t}{2} \cosh t} 
\sin\frac{\beta t}{\pi}\right] \,\,\,.
\label{integralU}
\EN 
The angle $\alpha$ is a positive quantity, related to the coupling
constant $\lambda$ of the model by 
\EQ
\alpha = \frac{1}{4 \pi} \frac{\lambda^2}{1 + \frac{\lambda^2}{4 \pi}}
\,\,\, .
\label{relationcouplings}
\EN 
This equation implies that the SShG is a quantum field theory 
invariant under the strong-weak duality 
\EQ
\lambda \rightarrow \frac{4\pi}{\lambda} \,\,\,.
\label{duality}
\EN 
It is easy to see that such $S$-matrix fulfills all constraints of 
section 2. The prefactor of the $S$-matrix admits the following infinite 
product representation 
\begin{eqnarray} 
&& Y(\beta)= \frac{1}
{
\Gamma\left(-i\frac{\beta}{2 \pi}\right)
\Gamma\left(-i\frac{\beta}{2 \pi}+\frac{1}{2}\right)
\Gamma\left(\frac{1}{2} +i\frac{\beta}{2 \pi}\right)
\Gamma\left(1+i\frac{\beta}{2 \pi}\right)} \nonumber \\
&& \,\,\,\times \prod_{k=0}^{\infty} 
\frac{
\Gamma\left(k + \alpha + \frac{1}{2} +i\frac{\beta}{2 \pi}\right)
\Gamma\left(k + \frac{3}{2} - \alpha +i\frac{\beta}{2 \pi}\right)
\Gamma^2\left(k+1+i\frac{\beta}{2 \pi}\right)}
{\Gamma\left(k + \alpha + \frac{1}{2} -i\frac{\beta}{2 \pi}\right)
\Gamma\left(k + \frac{3}{2} - \alpha -i\frac{\beta}{2 \pi}\right)
\Gamma^2\left(k+1-i\frac{\beta}{2 \pi}\right)}
\label{gammafuncpref}\\
&& \,\,\,\times\prod_{k=0}^{\infty}
\frac{
\Gamma\left(k + \alpha -i\frac{\beta}{2 \pi}\right)
\Gamma\left(k + 1 - \alpha -i\frac{\beta}{2 \pi}\right)
\Gamma^2\left(k+\frac{1}{2}-i\frac{\beta}{2 \pi}\right)}
{\Gamma\left(k + \alpha + 1 +i\frac{\beta}{2 \pi}\right)
\Gamma\left(k+2 - \alpha + i\frac{\beta}{2 \pi}\right)
\Gamma^2\left(k+\frac{3}{2}+i\frac{\beta}{2 \pi}\right)}
\nonumber
\end{eqnarray}
from which one can explicitly sees that this $S$-matrix does not 
have poles into the physical strip. It is now interesting to 
analyse several analytic continuations of the above $S$-matrix 
in the parameter $\alpha$. 

\subsection{The analytic continuation $\alpha \rightarrow -\alpha$ 
and Schoutens's model} 

Under this analytic continuation, the SShG model goes into the 
Super Sine-Gordon model and the $S$-matrix (\ref{SmatSSHG}) describes 
in this case the scattering of the lowest breather states of the 
latter model. The explicit expression is given by
\EQ 
S(\beta) = \hat Y(\beta)\,
\left(\begin{array}{cccc} 
1+\frac{2 i \sin\pi\alpha}{\sinh\beta} & 
\frac{\sin\pi\alpha}{\cosh\frac{\beta}{2}} 
& 0 & 0 \\ 
\frac{\sin\pi\alpha}{\cosh\frac{\beta}{2}} & 
-1 +\frac{2 i \sin\pi\alpha}{\sinh\beta} & 0 & 0 \\
0 & 0 &\frac{i\sin\pi\alpha}{\sinh\frac{\beta}{2}} & 1 \\
0 & 0 & 1 & 
\frac{i\sin\pi\alpha}{\sinh\frac{\beta}{2}}
\end{array}
\right) 
\label{SmatSSG}\\
\EN 
where 
\EQ
\hat Y(\beta) = \frac{\sinh\frac{\beta}{2}}{\sinh\frac{\beta}{2} - 
i \sin\pi\alpha} \,\hat U(\beta) \,\,\,,
\label{hatY}
\EN
and 
\EQ
\hat U(\beta) = U(\beta) \,
\left(\frac{\sinh\frac{\beta}{2} - i \sin\pi\alpha}
{\sinh\frac{\beta}{2} + i \sin\pi\alpha}\right) \, 
\left(\frac{\sinh\beta + i \sin(2\pi\alpha)}
{\sinh\beta - i \sin(2\pi\alpha} \right) \,\,\,.
\label{hatU}
\EN 
Notice that at the particular value $\alpha=\pi/3$ we can 
consistently truncate the theory at the Super Sine-Gordon 
breather sector only\footnote{The general class of models 
considered by Schoutens is obtained by taking 
$\alpha = \pi/(2 N +1)$, $N=1,2,\ldots$ and they correspond to 
the supersymmetric deformation of the non-unitary minimal superconformal 
models with central charge $C=-3N (4N+3)/(2N+2)$. For
simplicity only the first is examined, the detailed 
discussion of the others requires the application of the bootstrap 
equations to their $S$-matrix.}. At this value, 
the pole at $\beta =2 \pi i/3$ of the $S$-matrix can be regarded 
as due to the bosonic and fermionic one-particle states $\mid
b(\beta)\rangle $ and $\mid f(\beta)\rangle$. These particles are 
therefore bound states of themselves, in the channels 
\begin{eqnarray*}
&& b b \rightarrow b \rightarrow bb \,\,\,;\\
&& f f \rightarrow b \rightarrow f f \,\,\,,
\end{eqnarray*}
for the bosonic particle $b$, and in the channels 
\begin{eqnarray*}
&& b f \rightarrow f \rightarrow f b\,\,\,;\\
&& f b \rightarrow f \rightarrow b f\,\,\, ,
\end{eqnarray*}
for the fermionic particle $f$. There is of course a price to pay for 
this truncation: this means that the residues of the $S$-matrix at 
these poles will be purely imaginary. Such a model 
therefore would be the supersymmetric analogous of the Yang-Lee model 
for the ordinary Sine-Gordon model \cite{YL}. It was considered 
originally by Schoutens \cite{Schoutens} and it has been identified 
with the off-critical supersymmetric deformation of the non-unitary 
superconformal minimal model with central charge $C=-21/4$. The 
residues of the $S$-matrix are given by (Figure 2)
\begin{eqnarray}
&& \Gamma_{bb}^b = i \,\sqrt{3 \sqrt{3}}\, \kappa \,\,\,; \nonumber\\
&& \Gamma_{ff}^b = i \,\sqrt{3} \,\kappa \,\,\,;
\label{couplings}\\
&& \Gamma_{bf}^f = i \,\sqrt{3} \,\kappa \,\,\,;\nonumber \\
&& \Gamma_{fb}^f = i \,\sqrt{3} \,\kappa \,\,\,,\nonumber 
\end{eqnarray}
where 
\EQ
\kappa = \exp\left[-\frac{1}{2} \int_0^{\infty} \frac{dt}{t} 
\frac{\sinh\frac{t}{3}\,\sinh^2\frac{2t}{3}}{\cosh^2\frac{t}{2} \cosh t} 
\right] = \,0.7941001...
\label{kappa}
\EN 
 
We will come back to this model for the discussion of its 
Form Factors.

\subsection{The Roaming Models} 

The $S$ matrix (\ref{SmatSSHG}) of the SShG model has zeros in 
the physical strip located 
at $\alpha_1=i\pi\alpha$ and $\alpha_2=i\pi(1-\alpha)$. By varying the 
coupling constant $\lambda$, they move along the imaginary axis and 
they finally meet at the point $i\pi/2$, at the self-dual value of the 
coupling constant $\lambda^2 = 4\pi$. If we further increase the 
value of the coupling constant, they simply swap positions. 
But there is a more interesting possibility: as first proposed by 
Al. Zamolodchikov for the analogous case of the ordinary Sinh-Gordon 
model \cite{roaming}, once the two zeros meet at $i\pi$, they can  
enter the physical strip by taking complex values of the 
coupling constant (Figure 3). In this way, the location of the two 
zeros are given by 
\EQ
\alpha_{\pm} = \frac{1}{2} \pm i \alpha_0 \,\,\, .
\label{complexzeros}
\EN
From the analytic $S$-matrix theory, the existence of complex zeros 
in the physical strip implies the presence of resonances 
in the system. By analysing the finite-size behaviour of the theory by
means of the Thermodynamic Bethe Ansatz \cite{Martins}, the interesting 
result is that the net effect of these resonances consists in an infinite 
cascade of massless Renormalization Group flows generated by the 
Neveu-Schwartz fields ${\cal N}_{1,3}^m$ and passing through all 
minimal superconformal model ${\cal SM}_m$ with non-zero Witten index 
(see Figure 1). As discussed in the previous section, the ending point of
this infinite-nested RG flow should describe the ${\cal N}_{1,3}$ deformation 
of the superconformal model ${\cal SM}_{4}$. Is this really the case? By
taking  the limit $\alpha \rightarrow i \infty$ into the $S$-matrix 
(\ref{SmatSSHG}), it is easy to see that it reduces to the simplest 
supersymmetric $S$-matrix analysed in Section 3, which in turn 
coincides with the one of the Sine-Gordon model at $\xi=2\pi$. Since 
this $S$-matrix describes a massive deformation of the $C=1$ model, 
in order to confirm the above roaming trajectory scenario 
the only thing that remains to check is the comparison of the anomalous 
dimension of deforming field. In the Sine-Gordon model at $\xi=2\pi$, 
the anomalous dimension of the deforming field was determined to be 
$\Delta =2/3$, which is indeed the conformal dimension of the top 
component of the superfield ${\cal N}_{1,3}$ in the model ${\cal SM}_4$ !
In the light of this result, it is now clear why in the roaming limit 
the value which is actually selected is $\xi=2\pi$ among 
all possible values of the coupling constant of Sine-Gordon model. 

In the next section we will see that the identity between the $S$-matrix  
of the two models also implies an identity between the Form Factors of 
the two theories. 

\resection{Form Factors of the Trace Operators of the SShG Model}

For integrable quantum field theories, the knowledge of the
$S$-matrix is very often the starting point for a complete 
solution of quantum field dynamics in terms of an explicit
construction of the correlation functions of all fields of 
the theory. This result can be obtained by computing first 
the matrix elements of the operators on the asymptotic 
states (the so-called Form Factors) \cite{KW,Smirnov} and then 
inserting them into the the spectral representation of the 
correlators. For instance, in the case of the two-point 
correlation function of a generic operator ${\cal O}(z,\bar z)$ 
we have 
\EQ
{\cal G}(z,\bar z) = 
\langle 0 \mid {\cal O}(z,\bar z) {\cal O}(0,0)\mid 0\rangle = 
\int_0^{\infty} da^2 \,\rho(a^2) \, K_0(a \sqrt{z\bar z}) \,\,\, ,
\label{spectralrep}
\EN 
where $K_0(x)$ is the usual Bessel function. The spectral density 
$\rho(a^2)$ is given in this case by 
\begin{eqnarray}
\rho(a^2) &=& \sum_{n=0}^{\infty} 
\int 
\frac{d\beta_1}{2\pi}
\cdots 
\frac{d\beta_n}{2\pi}
\,\delta(a-\sum_i^n m \cosh\beta_i) \,\,\delta(\sum_i^n \sinh\beta_i) 
\times \label{rho} \\
&& \,\,\,\,\,\,\,\,\,\,\,\,\,\,
\mid \langle 0\mid {\cal O}(0,0)\mid A_1(\beta_1)\ldots
A_n(\beta_n) \rangle\mid^2 \,\,\, . \nonumber 
\end{eqnarray} 
The Form-Factor approach has proved to be extremely successful 
for theories with scalar $S$-matrix, leading to an explicit 
solution of models of statistical mechanics interest such as the 
Ising model \cite{KW,JLG,YZ}, the Yang-Lee model \cite{ZamYL} 
or quantum field theories defined by a lagrangian, like 
the Sinh-Gordon model \cite{KM,FMS,Korepin}. On the contrary, 
for theories with a non-scalar $S$-matrix the functional
equations satisfied by the Form Factors are generally quite 
difficult to tackle and a part from the Sine-Gordon model or theories 
which can be brought back to it \cite{KW,Smirnov,KF}, there is presently no 
mathematical technique available for solving them in their 
full generality. Also in this case, however, the situation is not 
as impractible as it might seem at first sight. The reason 
consists in the fast convergent behaviour of the spectral 
representation series which approximates the correlation functions with a 
high level of accuracy even if truncated at the first available 
matrix elements \cite{IMMF,YZ,ZamYL,CMpol,DM,CD,Saleur}.  
In the light of this fact, in this section we will compute 
the lowest matrix elements of two of the most important operators 
of the theory, namely the trace
operators  $\Theta(z,\bar z)$, $\chi(z,\bar z)$ and 
$\barr\chi(z,\bar z)$ of the supersymmetric stress-energy tensor 
of the SShG model. For the operator $\Theta(0,0)$, they 
are given by\footnote{They depend on the difference of rapidities 
$\beta =\beta_1-\beta_2$ since $\Theta(0,0)$ is a scalar operator.} 
\EQ
\begin{array}{l}
F_{bb}^{\Theta}(\beta) = \langle 0\mid \Theta(0,0)
\mid b(\beta_1) b(\beta_2)\rangle \,\,\,;\\
F_{ff}^{\Theta}(\beta) 
=\langle 0\mid \Theta(0) \mid f(\beta_1) f(\beta_2)\rangle \,\,\, ,
\end{array}
\label{FFTheta}
\EN 
whereas for the operators $\chi(0,0)$ we have instead
\EQ
\begin{array}{l} 
F_{bf}^{\chi}(\beta_1,\beta_2) = \langle 0\mid \chi(0,0)
\mid b(\beta_1) f(\beta_2)\rangle \,\,\,;\\
F_{fb}^{\chi}(\beta_1,\beta_2) = \langle 0\mid \chi(0,0)
\mid f(\beta_1) b(\beta_2)\rangle \,\,\,,
\end{array}
\label{FFchi}
\EN
(with an analogous result for the lowest Form Factors of the 
operator $\barr\chi(0,0)$). Since the operators $\Theta$, $\chi$ 
and $\barr\chi$ are related each other by supersymmetry, as 
consequence of eqs.\,(\ref{susytransftrace}) we have 
\begin{eqnarray}
&& F_{bb}^{\Theta}(\beta) = \,\,\,\omega \,\left(e^{\beta_1/2} \,
F_{fb}^{\chi} + e^{\beta_2/2}\,F_{bf}^{\chi}\right) \,\,\,;
\nonumber \\
&& F_{ff}^{\Theta}(\beta) = -\barr\omega \,\left(e^{\beta_2/2} \,
F_{fb}^{\chi} - e^{\beta_1/2}\,F_{bf}^{\chi}\right) \,\,\,;
\\
&& F_{bb}^{\Theta}(\beta) = \,\,\,\barr\omega \,\left(e^{-\beta_1/2} \,
F_{fb}^{\barr\chi} + e^{-\beta_2/2}\,F_{bf}^{\barr\chi}\right) \,\,\,;
\nonumber \\
&& F_{ff}^{\Theta}(\beta) = -\omega \,\left(e^{-\beta_2/2} \,
F_{fb}^{\barr\chi} - e^{-\beta_1/2}\,F_{bf}^{\barr\chi}\right) \,\,\,.
\nonumber 
\end{eqnarray}
It is therefore sufficient to compute the two-particle Form Factors 
of the operator $\Theta(z,\bar z)$ for determining those of 
$\chi(z,\bar z)$ and $\barr\chi(z,\bar z)$. Let us discuss the functional 
equations satisfied by $F_{bb}^{\Theta}(\beta)$ and 
$F_{ff}^{\Theta}(\beta)$.

The first set of equations (called the unitarity equations) rules the 
monodromy properties of the matrix elements as dictated by the 
$S$-matrix amplitudes  
\begin{eqnarray}
&& F_{bb}^{\Theta}(\beta) = S_{bb}^{bb}(\beta)\,
F_{bb}^{\Theta}(-\beta) + S_{bb}^{ff}(\beta) \,
F_{ff}^{\Theta}(-\beta) \,\,\,;
\label{unitarityFF} \\
&& F_{ff}^{\Theta}(\beta) = S_{ff}^{bb}(\beta)\,
F_{bb}^{\Theta}(-\beta) + S_{ff}^{ff}(\beta) \,
F_{ff}^{\Theta}(-\beta) \,\,\,,\nonumber
\end{eqnarray}
where $S_{bb}^{bb}(\beta)$ is the scattering amplitude of two 
bosons into two bosons and similarly for the other amplitudes. 

The second set of equations (called crossing equations) 
express the locality of the operator $\Theta(z,\bar z)$ 
\EQ
\begin{array}{l}
F_{bb}^{\Theta}(\beta + 2\pi i) = 
F_{bb}^{\Theta}(-\beta) \,\,\,;\\
F_{ff}^{\Theta}(\beta + 2\pi i) = 
F_{ff}^{\Theta}(-\beta) \,\,\,.
\end{array}
\label{crossingFF}
\EN 

One may be inclined to solve the coupled monodromy-crossing equations 
by initially diagonalising the $S$-matrix \cite{KW}. However, this 
method does not work in this case for a series of reasons, both of 
mathematical and physical origin. To simplify the notation, let us 
use in the following  
\begin{eqnarray*} 
&& a \equiv \sin\pi\alpha \,\,\, ;\\
&& s \equiv \sinh\frac{\beta}{2} \,\,\, ; \\
&& c \equiv \cosh\frac{\beta}{2} \,\,\, . 
\end{eqnarray*}
Observe that the eigenvalues of the $S$-matrix of the SShG model 
in the $F=1$ sector are given by 
\EQ
\zeta_{\pm} = -\frac{i a}{s c} 
\pm 
\sqrt{1 + \left(\frac{a}{c}\right)^2}
\,\,\, ,
\label{sqrt}
\EN 
with the corresponding eigenvectors  
\begin{eqnarray}
&& \mid \zeta_+ \rangle = N_+ \left[\frac{a}{c}
\,\mid b(\beta_1) b(\beta_2)\rangle + 
\left(1 - \sqrt{1 + \left(\frac{a}{c}\right)^2}\right)
\,\mid f(\beta_1) f(\beta_2)\rangle \right] \,\,\,;
\label{wreigen}\\ 
&& \mid \zeta_-\rangle = N_- \left[\frac{a}{c}
\,\mid b(\beta_1) b(\beta_2)\rangle + 
\left(1 + \sqrt{1 + \left(\frac{a}{c}\right)^2}\right) 
\,\mid f(\beta_1) f(\beta_2)\rangle \right] \,\,\, , 
\end{eqnarray}
and the normalization constants given by  
\[
N_{\pm} = \frac{1}{\sqrt{2 
\left(1 + \left(\frac{a}{c}\right)^2 
\mp \sqrt{1 + \left(\frac{a}{c}\right)^2}\right)}} \,\,\,.
\] 
From a mathematical point of view, the branch cuts present in the
eigenvalues (\ref{sqrt}) make it impossible to find their exponential
integral representations -- a step which is usually rather crucial in 
obtaining the corresponding Form Factor \cite{KW}. A more serious 
aspect, however, is the fact that the eigenvectors (\ref{wreigen}) 
do not have any satisfactory properties under the crossing transformation 
$\beta \rightarrow \beta + 2\pi i$. From a physical point of view, 
the origin of all these troubles is the different scattering 
property of the channel $b b \rightarrow b b $ with respect to 
the channel $ ff \rightarrow f f$, which does not permit, in this case,  
to assign a reasonable physical meaning to the states which diagonalise 
the $S$-matrix. 

The determination of the Form Factors of the operator $\Theta(z,\bar z)$ 
must pass through a different route. The way that we will 
proceed is to introduce two auxiliary functions $F_{\pm}(\beta)$ by 
considering the scattering theory in the $F=-1$ sector and to use 
them as building blocks for constructing the matrix elements 
$F_{bb}^{\Theta}(\beta)$ and $F_{ff}^{\Theta}(\beta)$. 

\subsection{Auxiliary problem: two-particle FF in the $F=-1$ sector}

Let us look for the eigenvalues of the $S$-matrix in the $F=-1$ sector. 
They are given by 
\begin{eqnarray}
&& \lambda_+ =  
\frac{s - i a}{s+ i a} \,U(\beta) \,\,\,;
\label{eigenvaluesS}
\\
&& \lambda_- = -U(\beta) \,\,\, ,\nonumber 
\end{eqnarray}
with the relative eigenvectors given by 
\begin{eqnarray} 
&& \mid \lambda_+ \rangle 
= \frac{1}{\sqrt{2}}
\left( \mid b(\beta_1) f(\beta_2)\rangle + f(\beta_1) b(\beta_2)\rangle
\right) \,\,\,;
\label{eigenvectorS}
\\
&& \mid \lambda_- \rangle = 
\frac{1}{\sqrt{2}}
\left( \mid f(\beta_1) b(\beta_2)\rangle - b(\beta_1) f(\beta_2)\rangle
\right) \,\,\,.\nonumber
\end{eqnarray}
The states in this sector are necessarily coupled to operators 
with a non-zero fermionic quantum number. For the purpose of 
obtaining the auxiliary functions $F_{\pm}(\beta)$ which will 
be used as building blocks to construct $F_{bb}^{\Theta}(\beta)$ 
and $F_{ff}^{\Theta}(\beta)$, it is sufficient to consider the coupling 
of the eigenstates (\ref{eigenvectorS}) to a fictitious scalar operator 
$\Lambda(0)$ but with a non-zero fermionic quantum number. 
Let us denote the corresponding matrix elements as 
\EQ
\begin{array}{l}
F_{+} (\beta) \equiv 
\langle 0 \mid \Lambda(0)\mid \lambda_+(\beta_1,\beta_2)\rangle 
\,\,\,;\\
F_{-} (\beta) \equiv \langle 0 \mid \Lambda(0)\mid 
\lambda_-(\beta_1,\beta_2) \rangle 
\,\,\,.
\end{array}
\label{two-particleaux}
\EN 
Under the condition of unitarity, these matrix elements satisy the 
equations 
\begin{eqnarray}
&& F_+(\beta) = \frac{s - i a}{s + i a} \,U(\beta) \,F_+(-\beta)\,\,\,;
\label{monodromy} 
\\
&& F_-(\beta) = - U(\beta) \,F_-(-\beta) \,\,\,.\nonumber 
\end{eqnarray}
The crossing properties of the above matrix elements is more subtle. 
Let us consider, in fact, what happens to $F_+(\beta)$ 
and to $F_-(\beta)$ in the analytic continuation 
$\beta \rightarrow \beta + 2 \pi i$, i.e. when the first
particle in each of the two states entering the eigenvectors 
(\ref{eigenvectorS}) goes around the operator $\Lambda(0)$. 
Since this operator has a fermionic quantum number, if the particle 
which goes around the operator is also a fermion we get an extra 
phase $(-1)$, otherwise nothing, and therefore 
\EQ
\begin{array}{llc}
\langle 0 \mid \Lambda(0) \left( 
\mid b(\beta_1) f(\beta_2)\rangle + \mid f(\beta_1) b(\beta_2)\rangle
\right) &\longrightarrow & -   
\langle 0 \mid \Lambda(0) \left( 
\mid b(\beta_2) f(\beta_1)\rangle - \mid f(\beta_2) b(\beta_1)\rangle
\right) \\
\langle 0 \mid \Lambda(0) \left( 
\mid f(\beta_1) b(\beta_2)\rangle - \mid b(\beta_1) f(\beta_2)\rangle
\right) & \longrightarrow &    
- \langle 0 \mid \Lambda(0) \left( 
\mid b(\beta_2) f(\beta_1)\rangle + \mid f(\beta_2) b(\beta_1)\rangle
\right) 
\end{array}
\label{crossing1}
\EN 
i.e. under crossing the two matrix elements mix each other, 
\EQ
\begin{array}{l}
F_+(\beta + 2\pi i) = - F_-(-\beta) \,\,\,;\\
F_-(\beta + 2\pi i) = - F_+(-\beta) \,\,\,. 
\end{array}
\label{crossing2}
\EN 
In order to solve the coupled system of crossing-unitarity equations 
(\ref{monodromy}) and (\ref{crossing2}), it is convenient to separate
the problem into two steps. The first step consists in finding a 
function $G(\beta)$ which solves the monodromy equation involving 
only the function $U(\beta)$, with the usual crossing property, i.e. 
\EQ
\begin{array}{l} 
G(\beta) = U(\beta) G(-\beta) \,\,\,;\\
G(\beta + 2\pi i) = G(-\beta) \,\,\, ,
\end{array}
\label{unicroG}
\EN 
The explicit expression of the function $G(\beta)$ can be found in 
Appendix A. 

The second step consists instead in finding two functions 
$f_+(\beta)$ and $f_-(\beta)$ which solve the functional equations 
\begin{eqnarray}
&& f_+(\beta) = \frac{s - i a}{s + i a} \,f_+(-\beta)\,\,\,;
\nonumber \\
&& f_-(\beta) = - \,f_-(-\beta) \,\,\,; 
\label{monodromyy} \\
&& f_+(\beta + 2\pi i) = - f_-(-\beta) \,\,\,;\nonumber \\
&& f_-(\beta + 2\pi i) = - f_+(-\beta) \,\,\,. \nonumber 
\end{eqnarray}
To this aim, let us write initially the integral representation of the
eigenvalues $\lambda_+(\beta)$  
\EQ
\frac{s  - i a}{s+ i a} = 
\exp\left[-4 i \int_0^{\infty} \frac{dt}{t} 
\frac{\sinh\alpha t \,\sinh(1-\alpha)t}{\cosh t} \sin\frac{\beta t}{\pi}
\right] \,\,\,.  
\label{integralhalf}  
\EN 
The minimal solutions of (\ref{monodromyy}) are then given by 
\begin{eqnarray}
&& f_+(\beta) = \cosh\frac{\beta}{4} \,H_+(\beta) \,\,\,;
\label{solutionaux}\\
&& f_-(\beta) = i\sinh\frac{\beta}{4} \,H_-(\beta) \,\,\,,\nonumber 
\end{eqnarray}
where 
\begin{eqnarray}
&& H_+(\beta) = \exp\left[4 \int_0^{\infty} \frac{dt}{t} 
\frac{\sinh\alpha t \,\sinh (1-\alpha)t}{\cosh t \sinh 2t} 
\sin^2\left(\frac{\beta - 2\pi i}{2\pi}\right)t \right] \,\,\,;
\label{H+-} \\
&& H_-(\beta) = 
\exp\left[4 \int_0^{\infty} \frac{dt}{t} 
\frac{\sinh\alpha t \,\sinh (1-\alpha)t}{\cosh t \sinh 2t} 
\sin^2\frac{\beta}{2\pi}t \right] \,\,\,. \nonumber 
\end{eqnarray}
The function $H_-(\beta)$ admits the equivalent representation 
\EQ
H_-(\beta) = \prod_{k=0}^{\infty} 
\left | 
\frac
{
\Gamma\left(k+\frac{3}{2} + i \frac{\beta}{4 \pi}\right)
\Gamma\left(k+1-\frac{\alpha}{2} + i \frac{\beta}{4 \pi}\right)
\Gamma\left(k+\frac{1}{2} +\frac{\alpha}{2}+ i \frac{\beta}{4 \pi}
\right)}
{
\Gamma\left(k+\frac{1}{2} + i \frac{\beta}{4 \pi}\right)
\Gamma\left(k+\frac{3}{2}-\frac{\alpha}{2} + i \frac{\beta}{4 \pi}\right)
\Gamma\left(k+1+\frac{\alpha}{2} + i \frac{\beta}{4 \pi}
\right)}
\right |^2
\label{gammah}
\EN 
or the mixed one, which is more useful for numerical calculation 
\begin{eqnarray}
&& H_-(\beta) = 
\prod_{k=0}^{N-1} 
\left[
\frac
{
\left(1 +\left(\frac{\beta}{\pi(4 k +2)}\right)^2\right)
\left(1 +\left(\frac{\beta}{\pi(4 k +4+2\alpha)}\right)^2\right)
\left(1 +\left(\frac{\beta}{\pi(4 k +6+2\alpha)}\right)^2\right)}
{
\left(1 +\left(\frac{\beta}{\pi(4 k +6)}\right)^2\right)
\left(1 +\left(\frac{\beta}{\pi(4 k +2+2\alpha)}\right)^2\right)
\left(1 +\left(\frac{\beta}{\pi(4 k +4-2\alpha)}\right)^2\right)}
\right]^{k+1} \times \nonumber \\
&& \times \exp\left[4 \int_0^{\infty} \frac{dt}{t} 
\frac{\sinh\alpha t \,\sinh (1-\alpha)t}{\cosh t \sinh 2t}\,
\left(N+1 - N e^{-4 t}\right) e^{-4 N t} \, 
\sin^2\frac{\beta}{2\pi}t \right] \,\,\,.  
\end{eqnarray}
Since the functions $H_{\pm}(\beta)$ are related each other by the 
equation $H_{+}(\beta + 2 \pi i) = H_-(\beta)$, the corresponding 
infinite product or mixed representations of $H_+(\beta)$ follow easily.  
For large values of $\beta$, $H_{+}(\beta)$ and $H_-(\beta)$ go 
the same constant, given by 
\EQ
\lim_{\beta\rightarrow\infty} H_{\pm}(\beta) = 
\exp\left[2 \int_0^{\infty} \frac{dt}{t} 
\frac{\sinh\alpha t \,\sinh (1-\alpha)t}{\cosh t \sinh 2t} 
\right] \,\,\,. \nonumber 
\label{asymconst}
\EN 
So, summarising the results of this section, the final 
expressions of the functions $F_+(\beta)$ and $F_-(\beta)$ are 
given by 
\EQ
\begin{array}{l}
F_+(\beta) = G(\beta)\,f_+(\beta) \,\,\,;\\
F_-(\beta) = G(\beta)\,f_-(\beta) \,\,\, ,
\end{array}
\label{finalF}
\EN 
with $G(\beta)$ given in eq.(\ref{integralG}) of Appendix A and 
$f_{\pm}(\beta)$ given in eq.\,(\ref{solutionaux}). 

\subsection{Two-particle FF of the operator $\Theta$}

Let us look for the two-particle Form Factors of the trace 
$\Theta$ of the stress-energy tensor as linear combination of 
the two functions $F_+(\beta)$ and $F_-(\beta)$ above determined, 
i.e. 
\EQ
\begin{array}{l}
F_{bb}(\beta) = {\cal A}(\beta) F_+(\beta) + {\cal B}(\beta) F_-(\beta) 
\,\,\,;\\
F_{ff}(\beta) = {\cal C}(\beta) F_+(\beta) + {\cal D}(\beta) F_-(\beta) 
\,\,\, .
\end{array}
\label{linearcomb}
\EN 
Let us now plug them into eq.(\ref{unitarityFF}). By using the monodromy
equations satisfied by the $F_{\pm}(\beta)$ and by comparing the terms in
front of each of these functions, we obtain the following equations 
\begin{eqnarray}
&& {\cal A}(\beta) = \frac{s}{s -ia} \,\,\,\,\left[
\left(1-\frac{i a}{s c }\right) {\cal A}(-\beta) - 
\frac{a}{c} \,{\cal C}(-\beta) \right]\,\,\,;
\nonumber \\
&& {\cal B}(\beta) = -\frac{s}{s + i a}  \left[
\left(1-\frac{ i a}{ s c}\right) {\cal B}(-\beta) - 
\frac{a}{c} \,{\cal D}(-\beta) \right]\,\,\,;
\label{ABCD}\\
&& {\cal C}(\beta) = -\frac{s}{s - i a} \left[
\left(1+\frac{i a}{s c }\right) {\cal C}(-\beta) + 
\frac{a}{c} \,{\cal A}(-\beta) \right]\,\,\,;
\nonumber \\
&& {\cal D}(\beta) = \frac{s}{s + i a} \,\,\,\left[
\left(1+\frac{i a}{s c}\right) {\cal D}(-\beta) + 
\frac{a}{c} \,{\cal B}(-\beta) \right]\,\,\,. 
\nonumber 
\end{eqnarray}
By taking the free limit $\alpha \rightarrow 0$ in the above equations, 
it is easy to see that  
\EQ
\begin{array}{l}
{\cal A}(\beta) = {\cal A}(-\beta) \,\,\,;\\
{\cal B}(\beta) = -{\cal B}(-\beta) \,\,\,;\\
{\cal C}(\beta) = -{\cal C}(-\beta) \,\,\,;\\
{\cal D}(\beta) = {\cal D}(-\beta) \,\,\,.
\end{array}
\label{parityABCD}
\EN
Let us assume that the above parity properties of 
the functions ${\cal A},\ldots, {\cal D}$ are still valid for 
a non-zero value of the coupling constant $\alpha$ (after all, 
the coupling constant dependence of the matrix elements $F_{bb}$ 
and $F_{ff}$ should be already included into the functions 
$F{\pm}(\beta)$). Under this hypothesis, eqs.(\ref{ABCD})
provide the relationships 
\begin{eqnarray}
&& {\cal C}(\beta) = -i \,
\frac{c-1}{s}\, {\cal A}(\beta) \,\,\,;
\label{firststep} \\
&& {\cal D}(\beta) = i \,\frac{c+1}{s} \, {\cal B}(\beta) \,\,\,.
\nonumber
\end{eqnarray}
Hence, at this stage we have for the two-particle Form Factors of 
$\Theta(x)$ 
\begin{eqnarray}
&& F_{bb}(\beta) = {\cal A} \,F_{+} + {\cal B} \,F_- \,\,\,,
\label{secondstep} \\
&& F_{ff}(\beta) = -i\left[
\frac{c-1}{s}\, {\cal A} \,F_+ -
\frac{c+1}{s}\, {\cal B} \,F_- \right] \,\,\,.\nonumber 
\end{eqnarray}
In order to determine the two remaining functions ${\cal A}(\beta)$ and 
${\cal B}(\beta)$, let us consider once again the case 
$\alpha \rightarrow 0$ and let us impose the condition that 
in this limit the two-particle Form Factors reduce to their free limit 
\EQ
\begin{array}{l} 
F_{bb}(\beta) = \,\,\,2 \pi m^2 \,\,\,;\\
F_{ff}(\beta) = -2 \pi i \sinh\frac{\beta}{2} \,\,\,.
\end{array}
\label{freeFF}
\EN 
From this matching, the functions ${\cal A}$ and
${\cal B}$ are uniquely determined to be 
\begin{eqnarray}
&& {\cal A}(\beta) = 2\pi \cosh\frac{\beta}{4} \,\,\,,\\
&& {\cal B}(\beta) = 2\pi i \sinh\frac{\beta}{4} \,\,\,.\nonumber 
\end{eqnarray}
For a generic value of the coupling constant, the correct 
expressions of the two-particle Form Factors of the operator
$\Theta(0)$ can be otained by imposing their normalization 
$F_{bb}(i\pi) = F_{ff}(i\pi) = 2\pi m^2$ and their final form 
are given by 
\begin{eqnarray}
&&F_{bb}^{\Theta}(\beta) = 2\pi m^2\,\frac{\tilde F_{bb}(\beta)}
{\tilde F_{bb}(i\pi)} \,\,\,;
\label{finalFF} \\
&&F_{ff}^{\Theta}(\beta) = 2\pi m^2\, 
\frac{\tilde F_{ff}(\beta)}{\tilde F_{ff}(i\pi)} \,\,\, ,\nonumber
\end{eqnarray}
where 
\begin{eqnarray}
&& \tilde F_{bb}(\beta) = \left[
\cosh^2\frac{\beta}{4} \,H_+(\beta) - 
\sinh^2\frac{\beta}{4} \,H_-(\beta)\right]\,G(\beta) \,\,\, ,
\label{finaltildeFF} \\
&& \tilde F_{ff}(\beta) =  
\sinh\frac{\beta}{2} \,
\left[H_+(\beta) + H_-(\beta)\right]\,G(\beta) \,\,\, .  \nonumber 
\end{eqnarray}
Notice that for large values of $\beta$, $F_{bb}^{\Theta}(\beta)$ 
tends to a constant whereas $F_{ff}^{\Theta}(\beta) \simeq e^{\beta/2}$, 
both behaviour in agreement with Weinberg's power counting theorem 
of the Feynman diagrams. 

While we postpone non trivial checks of the validity of (\ref{finalFF}) 
to the next sections, let us use the Form Factors (\ref{finalFF})  
to estimate the correlation function $C(r)= \langle \Theta(r)
\Theta(0)\rangle$ by means of formulas (\ref{spectralrep}) and (\ref{rho}). 
In the free limit, the correlator is simply expressed in terms of 
Bessel functions, 
\EQ
C(r) = m^4 \left(K_1^2(m r) + K_0^2(m r)\right)\,\,\,. 
\label{freecor}
\EN 
For a finite value of $\alpha$, a numerical integration of 
(\ref{spectralrep}) produces the graphs shown in Figure 4. As 
it was expected, in the ultraviolet limit the curve relative to a 
finite value of $\alpha$ is stepest than the curve relative to the 
free case whereas it decreases slower at large values of $mr$.  
This curve is expected to correctly capture the long distance behaviour 
of the correlator and to provide a reasonable estimate of their 
short distance singularity. However, for the exact estimation of the power
law singularity at the origin one would of course need the knowledge of 
all higher particle Form Factors.

\resection{$C$-theorem Sum Rule}

As mentioned in Section 4.1, the SShG model can be seen as a massive 
deformation of the superconformal model with the central charge 
$C=\frac{3}{2}$. While this fixed point rules the ultraviolet 
properties of the model, its large distance behaviour is controlled 
by a purely massive theory with $C=0$. The variation of the central 
charge in this RG flow is dictated by the $C$-theorem of Zamolodchikov 
\cite{Zamcth}, which we will discuss in more details in the next 
section in relation with its fermionic formulation. In this section, 
we are concerning with the integral version of the $C$-theorem 
\cite{Cardycth} in order to have non trivial checks of the validity of the 
Form Factors (\ref{finalFF}). In this formulation of the $c$-theorem, 
the variation of the central charge $\Delta C$ satisfies the sum rule 
\EQ
\Delta C = \frac{3}{4\pi} \int d^2x\,\mid x\mid^2\, 
\langle 0\mid\Theta(x) \Theta(0)\mid 0\rangle_{conn} \,= 
\int_0^{\infty} d\mu\, c(\mu)\,\, ,
\label{sumrule}
\EN 
where $c(\mu)$ is given by 
\begin{eqnarray}
& &c(\mu) = \frac{6}{\pi^2} \frac{1}{\mu^3} {\rm Im} \,G(p^2=-\mu^2) \,\,\,,
\nonumber \\
&& G(p^2) = \int d^2x e^{-i p x}  
\langle 0\mid\Theta(x) \Theta(0)\mid 0\rangle_{conn} \,\,\,.
\label{spectralcth}
\end{eqnarray}
Inserting a complete set of in-state into (\ref{spectralcth}), the 
spectral function $c(\mu)$ can be expressed as a sum on the FF's 
\begin{eqnarray}
c(\mu) &=& \frac{12}{\mu^3} \sum_{n=1}^{\infty} 
\int \frac{d\beta_1}{2\pi}
\cdots 
\frac{d\beta_n}{2\pi}
\,\delta(\mu -\sum_i^n m \cosh\beta_i) \,\,\delta(\sum_i^n m\sinh\beta_i) 
\times \label{cspect} \\
&& \,\,\,\,\,\,\,\,\,\,\,\,\,\,
\mid \langle 0\mid \Theta(0,0)\mid A_1(\beta_1)\ldots
A_n(\beta_n) \rangle\mid^2 \,\,\, . \nonumber 
\end{eqnarray} 
Since the term $\mid x\mid^2$ present in (\ref{sumrule}) suppresses 
the ultraviolet singularity of the two-point correlator of 
$\Theta$, the sum rule (\ref{sumrule}) is expected to be saturated 
by the first terms of the series (\ref{cspect}). For the SShG model 
the first approximation to the sum rule (\ref{sumrule}) is given by 
the contributions of the two-particle states 
\EQ
\Delta C^{(2)} = \frac{3}{8\pi^2 m^4} \int_0^{\infty} 
\frac{d\beta}{\cosh^4\beta} \,\left[
\mid F_{bb}^{\Theta}(2\beta)\mid^2 +
\mid F_{ff}^{\Theta}(2\beta)\mid^2 \right] \,\,\,.
\label{twocth}
\EN 
The numerical data relative to the above integral for different values 
of the coupling constant $\alpha$ is reported in Table 1. They are 
remarkably close to the theoretical value $\Delta C = \frac{3}{2}$, 
even for the largest possible value of the coupling constant, which 
is the self-dual point $\lambda = \sqrt{4\pi}$. In addition to 
this satisfactory check, a more interesting result is obtained by 
analysing the application of the $c$-theorem sum rule to models 
which are obtained as analytic continuation of the SShG.

\subsection{$C$-theorem Sum Rule for the Roaming Model} 

As we have seen in Section 5.2, by taking the analytic continuation 
\EQ
\alpha \rightarrow \frac{1}{2} + i \alpha_0 \,\,\,
\label{rrr}
\EN 
and the limit $\alpha_0 \rightarrow \infty$, the $S$-matrix of the 
SShG model is mapped into the $S$-matrix of the ordinary Sine-Gordon 
model at $\xi = 2\pi$. This identity between the two $S$-matrices is 
expected to extend to other properties of the two theories as well. There
are however some subtleties in this mapping. 

One subtlety is that the right hand side of the sum rule
(\ref{sumrule}) is a pure number ($\Delta C =\frac{3}{2}$ for the 
SShG model) -- a quantity which therefore seems to be completly 
insensitive to the variation of the coupling constant of the model. 
On the other hand, if the SShG model is mapped onto the Sine-Gordon model
in the limit $\alpha_0 \rightarrow \infty$ of the analytic continuation 
(\ref{rrr}), the sum rule should instead jump {\em discontinously} from the 
value $\Delta C = \frac{3}{2}$ to $\Delta C =1$, simply because 
the latter model is a massive deformation of the conformal theory 
with $C=1$. This is indeed what happens\footnote{We will not discuss 
further this statement since a detailed discussion of this aspect can be 
found in \cite{ADM}, where the analogous situation of the 
roaming limit mapping between the Sinh-Gordon model and the Ising model 
is analysed.}, the reason being the presence of an additional energy 
scale brought by the parameter $\alpha_0$ and the non-uniform 
convergence of the spectral series (\ref{cspect}). 

In the case of the SShG model, there is an additional point 
that however deserves to be carefully checked. Note that in the SShG 
model, the lowest approximation of the sum rule, eq.\,(\ref{twocth}), is 
expressed as sum of the moduli square of both $F_{bb}^{\Theta}$ 
and $F_{ff}^{\Theta}$. In the roaming limit, the original states 
$\mid b b\rangle$ and $\mid f f \rangle$ of the Sinh-Gordon model 
are mapped respectively onto the states $\mid S \barr S\rangle$ 
and $\mid\barr S S\rangle$ of the Sine-Gordon model. On the other hand, 
the trace $\Theta$ of the stress-energy tensor of the Sine-Gordon 
model only couples to the symmetric combination of the soliton states  
\EQ
\mid +\rangle = \frac{1}{\sqrt 2} \left(\mid S \barr S\rangle + 
\mid\barr S S\rangle\right) \,\,\,,
\label{symmetricSG}
\EN 
and not to each of them individually! How this apparent 
discrepancy can be settled?  

First let us consider how the two-particle Form Factor of the 
operator $\Theta$ is computed within the Sine-Gordon theory for  
a generic value of $\xi$ \cite{Smirnov}. The symmetry responsable for 
the coupling of the operator $\Theta$ to the combination 
(\ref{symmetricSG}) is the invariance of $\Theta$ under the charge 
conjugation. Hence, to compute the two-particle Form Factor 
\EQ
F_+^{SG}(\beta) = \langle 0 \mid \Theta(0)\mid +\rangle \,\,\, ,
\EN 
one 
needs to employ the corresponding eigenvalues of the $S$-matrix 
(\ref{sgsmatrix}) given by 
\EQ
\lambda_+^{SG}(\beta) = \frac
{\sinh\frac{\pi}{2\xi}(\beta + i \pi)}
{\sinh\frac{\pi}{2\xi}(\beta - i \pi)} \,\hat R(\beta) \,\,\,,
\label{eigSG}
\EN 
and solve the unitarity and crossing equations 
\EQ
\begin{array}{l}
F_+^{SG}(\beta) = \lambda_+(\beta) \,F_+^{SG}(-\beta) \,\,\,;\\
F_+^{SG}(\beta + 2\pi i) = F_+^{SG}(-\beta) \,\,\,.
\end{array}
\label{FF2SG}
\EN 
The solution of these equation gives us the two-particle Form Factor 
of $\Theta$ of the Sine-Gordon model
\EQ
F_+^{SG}(\beta) = -\frac{{\sqrt 2} \pi^2 m^2}{\xi \hat G(i\pi,\xi)} \,
\left[
\frac{\sinh\beta}{\sinh\frac{\pi}{2\xi}(\beta - i\pi)} 
\right] \,\hat G(\beta,\xi) \,\,\,,
\label{minSG}
\EN 
with 
\EQ
\hat G(\beta,\xi) = \exp\left[\int_0^{\infty}\frac{dt}{t} 
\frac{\sinh\frac{t}{2}\left(1-\frac{\xi}{\pi}\right)}
{\cosh\frac{t}{2} \sinh\frac{\xi t}{2 \pi} \sinh t} 
\sin^2\left(\frac{i\pi-\beta}{2\pi}\right)t\right] \,\,\,.
\label{GSG}
\EN 
Observe that for $\xi=2\pi$, the above expression (\ref{minSG}) can 
be written equivalently as 
\EQ
F_+^{SG}(\beta) = -\frac{{\sqrt 2} \pi m^2}{\hat G(i\pi,2\pi)}
 \,\sinh\frac{\beta}{2} \,
\left[\sinh\frac{\beta}{4} + i \cosh\frac{\beta}{4} \right] 
\,\hat G(\beta,2\pi) \,\,\,. 
\label{FFSG2pi}
\EN 
Once inserted into the sum rule (\ref{sumrule}), the result is 
$\Delta C^{(2)} = 0.9924...$, i.e. a saturation within few percent 
of the exact value $\Delta C =1$ relative to this case. 

Let us consider now the corresponding FF of the SShG model in the 
roaming limit, eq.\,(\ref{finalF}). For $\alpha_0 \rightarrow \infty$, 
the function $G(\beta)$ reduces precisely to $\hat G(\beta,2\pi)$ whereas 
for $H_{\pm}(\beta)$ we have 
\begin{eqnarray}
&& H_+(\beta) = \exp\left[2 \int_0^{\infty} \frac{dt}{t} 
\frac{\sin^2\left(\frac{\beta - 2\pi i}{2\pi}\right)t}
{\sinh 2t} \right] \,\,\,;
\label{H+-roa} \\
&& H_-(\beta) = 
\exp\left[2 \int_0^{\infty} \frac{dt}{t} 
\frac{\sin^2\frac{\beta}{2\pi}t}{\sinh 2 t} \right] \,\,\,. 
\nonumber 
\end{eqnarray}
By using the formula 
\[
\int_0^{\infty} \frac{dt}{t} e^{-p t} \sin^2\frac{a t}{2} \,=\,
\frac{1}{4} \ln \left[1 + \left(\frac{a}{p}\right)^2\right] \,\,\,, 
\label{use}
\] 
and the infinite product representation 
\[
\cosh x = \prod_{k=0}^{\infty} \left(1 + \frac{4 x^2}{(2 k+1)^2 \pi^2}
\right) \,\,\, ,
\label{cosh}
\]
it is easy to see that they can be simply expressed as 
\EQ
H_+(\beta) = -i \sinh\frac{\beta}{4} 
\,\,\,\,\, , 
\,\,\,\,\, 
H_-(\beta) = \cosh\frac{\beta}{4} \,\,\,.
\label{limitH}
\EN 
Hence, in the roaming limit the two matrix elements 
$F_{bb}^{\Theta}(\beta)$ and $F_{bb}^{\Theta}(\beta)$ become 
equal, with their common value given by 
\EQ
F_{bb}^{\Theta}(\beta) = F_{ff}^{\Theta}(\beta) = 
-\frac{{\sqrt 2} \pi m^2}{\hat G(i \pi,2\pi)} \, \sinh\frac{\beta}{2} 
\,\left[\sinh\frac{\beta}{4} + i \cosh\frac{\beta}{4}\right] \,
\hat G(\beta,2 \pi)  
\,\,\, ,
\label{finalroa}
\EN 
which coincides with the one of the Sine-Gordon model, 
eq.\,(\ref{FFSG2pi}). Therefore their contribution to the 
$c$-theorem sum rule is precisely the same as the Sine-Gordon 
model (in the case of the SShG model, eq.\,(\ref{twocth}),  
there is in fact a factor $\frac{1}{2}$ with respect to the 
Sine-Gordon model, which is cancelled by the equality of the 
two Form Factors $F_{bb}^{\Theta}$ and $F_{ff}^{\Theta}$). 

\subsection{$C$-theorem Sum Rule for Schoutens's Model} 
 
In the analytic continuation $\alpha \rightarrow -\alpha$, 
the $S$-matrix develops a pole in the physical strip located 
at $\beta = 2  \pi\alpha i$. As discussed in Section 5.1, the scattering 
theory for $\alpha = \frac{1}{3}$ admits a consistent interpretation 
in terms of a multiplet made of a boson and a fermion, which are 
bound states of themselves. For this model, the operator 
$\Theta(z,\bar z)$ has also a one-particle Form Factor 
$F_{b}^{\Theta}$ in virtue of the graph of Figure 5.    

In order to discuss the analytic continuation of the Form Factors 
$F_{bb}^{\Theta}(\beta)$ and $F_{ff}^{\Theta}(\beta)$ for 
$\alpha \rightarrow -\alpha$ and to calculate the matrix element 
$F_{b}^{\Theta}$, it is convenient to consider initially 
a different representation of the functions $f_{\pm}(\beta)$. This is 
given by 
\begin{eqnarray}
&& f_+(\beta) = \sinh\frac{\beta}{2} Z_+(\beta) \,\,\,;
\label{solutionaux2}\\
&& f_-(\beta) = -\sinh\frac{\beta}{2} Z_-(\beta) \,\,\,,\nonumber 
\end{eqnarray}
with 
\begin{eqnarray}
&& Z_+(\beta) = \exp\left[-2 \int_0^{\infty} \frac{dt}{t} 
\frac{\cosh(1-2\alpha)t}{\cosh t \sinh 2t} 
\sin^2\left(\frac{\beta - 2\pi i}{2\pi}\right)t \right] \,\,\,;
\label{Z+-} \\
&& Z_-(\beta) = 
\exp\left[-2 \int_0^{\infty} \frac{dt}{t} 
\frac{\cosh(1-2\alpha)t}{\cosh t \sinh 2t} 
\sin^2\frac{\beta}{2\pi}t \right] \,\,\,. \nonumber 
\end{eqnarray}
Their equivalence to the previous ones, eq.\,(\ref{solutionaux}) is 
easily established by comparing their infinite product representations
which for the function $Z_-(\beta)$ is given by 
\EQ
Z_-(\beta) = \prod_{k=0}^{\infty} \left|
\frac{
\Gamma\left(k+\frac{1}{2}+\frac{\alpha}{2} + i\frac{\beta}{4\pi}\right)
\Gamma\left(k+1-\frac{\alpha}{2} + i\frac{\beta}{4\pi}\right)}
{\Gamma\left(k+\frac{3}{2}-\frac{\alpha}{2} + i\frac{\beta}{4\pi}\right)
\Gamma\left(k+1+\frac{\alpha}{2} + i\frac{\beta}{4\pi}\right)}
\right|^2
\label{infiniteZ}
\,\,\,, 
\EN 
while the one of $Z_+(\beta)$ is simply given by the 
identity $Z_+(\beta+2\pi i) = Z_-(\beta)$. By using 
eq.\,(\ref{infiniteZ}) and the analogous for $Z_+(\beta)$, 
it is simply to see that under the analytic continuation 
$\alpha \rightarrow -\alpha$, the functions $f_{\pm}(\beta)$ 
change as follows 
\begin{eqnarray}
&& f_+(\beta) \rightarrow \hat f_+(\beta) = 
-\frac{2 \pi^2}{\cosh\frac{\beta}{2} - \cos\pi\alpha}\,Z_+^{-1}(\beta) 
\,\,\, ;\\
&& f_-(\beta) \rightarrow \hat f_-(\beta) = 
\frac{2 \pi^2}{\cosh\frac{\beta}{2} + \cos\pi\alpha}\,Z_-^{-1}(\beta) 
\,\,\, . \nonumber 
\end{eqnarray}
The analytic continuation of the two functions present therefore 
distinct physical properties: in fact, while the function 
$\hat f_+(\beta)$ develops a pole relative to the bound state, 
the other $\hat f_-(\beta)$ does not have any singularity in the 
physical strip.    

To compute the two-particle FF of $\Theta$ is not sufficient, 
however, to substitute the above functions $\hat f_{\pm}(\beta)$ into 
eq.\,(\ref{finalFF}). In fact, we must also take into account the extra 
terms present in $\hat U(\beta)$, eq.\,(\ref{hatU}). This is a 
simple task, though, because we only have to introduce the function 
$K(\beta) \equiv K_1(\beta) K_2(\beta)$, where $K_1(\beta)$ and 
$K_2(\beta)$ are solutions of the unitarity and crossing equations  
\begin{eqnarray}
&& K_1(\beta) = 
\left(\frac{\sinh\frac{\beta}{2} - i \sin\pi\alpha}
{\sinh\frac{\beta}{2} + i \sin\pi\alpha}\right) \,K_1(-\beta) \,\,\,;
\nonumber \\
&& K_1(\beta+2\pi i) = K_1(-\beta) \,\,\,;
\label{minK}\\
&& K_2(\beta) =   
\left(\frac{\sinh\beta + i \sin(2\pi\alpha)}
{\sinh\beta - i \sin(2\pi\alpha} \right)\,K_2(-\beta) \,\,\,; 
\nonumber\\
&& K_1(\beta + 2\pi i) = K_2(-\beta) \,\,\,.\nonumber 
\end{eqnarray}
Their mixed representation expressions are given by 
\begin{eqnarray}
&& K_1(\beta) = \prod_{k=0}^{N-1} 
\left[
\left(1 +\left(\frac{i\pi-\beta}
{4\pi(k+\frac{1}{4}+\frac{\alpha}{2})}\right)^2\right)
\left(1 +\left(\frac{i\pi-\beta}
{4\pi(k+\frac{3}{4}-\frac{\alpha}{2})}\right)^2\right) 
\right]^{-1} \times \nonumber \\
&& \,\, \times \,
\exp\left[-4 \int_0^{\infty} \frac{dt}{t} 
\frac{\cosh(1-2\alpha)t}{\sinh 2t}\, e^{-4 n t} \,
\sin^2\frac{(i\pi-\beta)}{2\pi}t\right] \,\,\,.
\label{K2mix}
\end{eqnarray}
and 
\begin{eqnarray}
&& K_2(\beta) = \prod_{k=0}^{N-1} 
\left[\frac{
\left(1 +\left(\frac{i\pi-\beta}
{2 \pi(k+\frac{1}{2}+\frac{\alpha}{2})}\right)^2\right)
\left(1 +\left(\frac{i\pi-\beta}{2\pi(k +1-\frac{\alpha}{2})}\right)^2\right)}
{
\left(1 +\left(\frac{i\pi-\beta}{2\pi(k +1+\frac{\alpha}{2})}\right)^2\right)
\left(1 +\left(\frac{i\pi-\beta}
{2\pi(k+\frac{3}{2}-\frac{\alpha}{2})}\right)^2\right)}
\right]^{k+1} \times \nonumber \\
&& \,\, \times \,
\exp\left[2 \int_0^{\infty} \frac{dt}{t} 
\frac{\cosh\frac{t}{2}(1-2\alpha)}{\sinh t \cosh\frac{t}{2}}
(N+1-N e^{-2 t}) e^{-2 n t} \,
\sin^2\frac{(i\pi-\beta)}{2\pi}t\right] \,\,\,.
\label{K1mix}
\end{eqnarray}

So, finally the two-particle FF of the operator $\Theta(z,\bar z)$ of 
the Schoutens's model are given by 
\begin{eqnarray}
&& F_{bb}^{\Theta}(\beta) = 2\pi m^2\, 
\frac{\tilde F_{bb}(\beta)}{\tilde F_{bb}(i\pi)} \,\,\,;
\label{finFFSCOU} \\
&& F_{ff}^{\Theta}(\beta) = 2\pi m^2 \,
\frac{\tilde F_{ff}(\beta)}{\tilde F_{ff}(i\pi)} \,\,\, , 
\nonumber 
\end{eqnarray}
where 
\begin{eqnarray}
&& 
\tilde F_{bb}(\beta) = \sinh\frac{\beta}{2} \,\left[\cosh\frac{\beta}{4}\, 
\frac{Z_+^{-1}(\beta)}
{\left(\cosh\frac{\beta}{2}-\cos\pi\alpha\right)} + 
i \sinh\frac{\beta}{4}\, 
\frac{Z_-^{-1}(\beta)}{\left(\cosh\frac{\beta}{2}+\cos\pi\alpha\right)} 
\right] \,G(\beta) K(\beta) \,\,\, ; 
\label{finalFFSco}  \nonumber\\
&& \tilde F_{ff}(\beta) = 
\sinh\frac{\beta}{2} \,
\left[\sinh\frac{\beta}{4} \,\frac{Z_+^{-1}(\beta)}
{\left(\cosh\frac{\beta}{2}
-\cos\pi\alpha\right)} - i \cosh\frac{\beta}{4}\,
\frac{Z_-^{-1}(\beta)}{\left(\cosh\frac{\beta}{2}+\cos\pi\alpha\right)}
\right]\, G(\beta) K(\beta) \,\,\, .  
\nonumber 
\end{eqnarray}
The one-particle Form Factor $F_{b}$ can now be equivalently 
obtained either from the residue equation on $F_{bb}^{\Theta}(\beta)$ 
or on $F_{ff}$
\begin{eqnarray}
&& -i \lim_{\beta\rightarrow \frac{2\pi i}{3}}
\left(\beta -\frac{2\pi i}{3}\right)\, F_{bb}^{\Theta}(\beta) 
= \Gamma_{bb}^b \,F_{b}^{\Theta} \,\,\, ;
\label{residueFF} \\
&& -i \lim_{\beta\rightarrow \frac{2\pi i}{3}}
\left(\beta -\frac{2\pi i}{3}\right)\, F_{ff}^{\Theta}(\beta) 
= \Gamma_{ff}^b \,F_{b}^{\Theta} \,\,\, ,
\nonumber
\end{eqnarray}
with the result 
\EQ
F_b^{\Theta} = -i \frac{\pi}{\kappa\sqrt{2\sqrt{3}}} \,
\frac{Z_+(i\pi)}{Z_+\left(\frac{2\pi i}{3}\right)}\,
G\left(\frac{2\pi i}{3}\right)\,K\left(\frac{2\pi i}{3}\right) = 
- 1.6719(3) \,i
\label{oneFFSco}
\EN 
The series of the sum rule has alternating sign, with the first 
contribution given by the one-particle Form Factor
\EQ
\Delta C^{(1)} = \frac{6}{\pi} (F_b^{\Theta})^2 = -5.3387(4) 
\label{onecth}
\EN 
This quantity differs for a $1.6\%$ from the theoretical value 
$\Delta C = -\frac{21}{4}=-5.25$. By also including the positive 
contribution of the two-particle FF, computed numerically  
\EQ
\Delta C^{(2)} = 0.09050(8) \,\,\, ,
\EN 
the estimate of the central charge of the model further improves, 
$C= -5.2482(4)$, with a difference from the exact value of just  
$0.033\%$.  

The Form Factors above determined for the trace operator $\Theta(x)$ 
can be used to estimate its two-point correlation function. 
The graph of this function is shown in Figure 6: note that this function 
diverges at the origin, in agreement with the positive conformal 
dimension $\Delta = 1/4$ of this operator, but it presents a non--monotonous 
behaviour for the alternating sign of its spectral series.

\resection{Fermionic Formulation of the $C$-theorem}

In the OPE (\ref{OPE}) of the fields $T(z)$ and $G(z)$ which are 
responsible for a superconformal symmetry, the central charge $C$ 
appears both in the short-distance singularity of $T(z) T(w)$ and 
$G(z) G(w)$. Therefore, going away from criticality by means of 
operators which preserve the supersymmetry of the critical point (but 
non necessarely its integrability), it should be possible to 
formulate a $C$-theorem for unitarity theories by looking at 
the fermionic sector. This is indeed the case, as proved in this 
section. 

Let us consider the correlation functions of the operators 
$G(z,\bar z)$ and $\chi(z,\bar z)$ away from criticality. By 
taking into account the scaling dimensions of the operators,  
these correlators can be parameterised as  
\begin{eqnarray}
\langle G(z,\bar z) G(0,0) \rangle \,=\,\frac{H(z\bar z)}{z^3} 
\,\,\,;\nonumber \\
\langle G(z,\bar z) \chi(0,0) \rangle \,=\,\frac{I(z\bar z)}{z^2 \bar z} 
\,\,\,;\label{corrGX}\\
\langle \chi(z,\bar z) \chi(0,0) \rangle \,=\,\frac{L(z\bar z)}{z \bar z^2} 
\,\,\,, \nonumber 
\end{eqnarray}
where the functions $H(z\bar z)$, $I(z\bar z)$ and $L(z\bar z)$ depend 
on the scalar variable $x\equiv z\bar z$. In virtue of the 
conservation law 
\EQ
\partial_{\bar z}\,G(z,\bar z) = \partial_z\,\chi(z,\bar z) \,\,\,,
\label{conGX}
\EN 
the above functions satisfy the following differential equations 
\EQ
\begin{array}{l}
\dot H - \dot I = - 2 I \,\,\,;\\
\dot I - \dot L = I - L \,\,\, ,
\end{array}
\label{differential}
\EN   
where $\dot H \equiv z\bar z \frac{d H}{dz\bar z}$ and similarly for the 
other functions. For the combination $F = H + I -L$ we have therefore 
\EQ
\dot F = - 2 L \,\,\,.
\label{difffinal}
\EN 
At the critical points, the quantity $F$ is proportional to the central 
charge $C$, since 
\EQ
\begin{array}{l}
H = \frac{2}{3}\,C \,\,\,;\\
I = L =0 \,\,\,.
\end{array}
\label{Fcritical}
\EN 
In order to conclude that the function $F$ decreases along the 
Renormalization Group flow and to derive the corresponding sum-rule, 
it is necessary to analyse the positivity of the function $L(z\bar z)$,
which is given by  
\EQ
L(z\bar z) = (z\bar z) \bar z \langle \chi(z,\bar z) \chi(0,0)\rangle
\label{LLL}\,\,\,.
\EN 
Although it seems a-priori difficult to prove the positivity of 
the above function for the explicit presence of the complex term 
$\bar z$, it will become evident that this term is precisely required 
for the proof. 

Let us start our analysis from the constraints provided by supersymmetry. 
By using eq.\,(\ref{susytransftrace}), we have 
\begin{eqnarray} 
&& \langle \Theta(z,\bar z) \Theta(0,0)\rangle = 
\langle \{\chi(z,\bar z),{\cal Q}\} 
\{\chi(0,0),{\cal Q}\}\rangle = \label{relationcorr}\\  
&& =\langle \chi(z,\bar z) {\cal Q}^2 \chi(0,0)\rangle = 
\langle \chi(z,\bar z) P_+ \chi(0,0)\rangle \,\,\,.\nonumber 
\end{eqnarray}
The above equations implies that the spectral function 
of the correlator $\langle \Theta(z,\bar z) \Theta(0,0)\rangle$ 
can be obtained from the one of  
$\langle \chi(z,\bar z) \chi(0,0)\rangle$ by multiplying 
each term of the multi-particle expansion for the right
momentum of the intermediate state. This is a positive 
quantity, since for a cluster of $n$-particles is given by 
\EQ
P_+ = m \,\left( e^{\beta_1} + e^{\beta_2} + 
\cdots e^{\beta_n} \right)\,\,\,.
\label{momentum}
\EN 
The positivity of the spectral series of 
$\langle \Theta(z,\bar z) \Theta(0,0)$ for unitarity theories 
implies therefore the positivity of the spectral series of 
$\langle \chi(z,\bar z) \chi(0,0)\rangle$ as well. 

Let us consider now the behaviour of the correlator 
$\langle \chi(z,\bar z) \chi(0,0)\rangle$ as a function 
of the variables $z$ and $\bar z$. Since (\ref{relationcorr}) 
can be written equivalently as 
\EQ
\langle \Theta(z,\bar z) \Theta(0,0)\rangle = 
\langle \chi(z,\bar z)\, \frac{M^2}{P_-} \, \chi(0,0) \rangle \,\,\,,
\label{massinser}
\EN 
where $M^2$ is the invariant mass square of the cluster of the 
intermediate particles, this implies that 
$\langle \chi(z,\bar z) \chi(0,0) \rangle$ can be expressed as 
\begin{eqnarray}
\langle \chi(z,\bar z)\chi(0,0) \rangle &=& 
\int \frac{d^2p}{2\pi} \,p_-\, \hat \chi(p^2) 
e^{-\frac{1}{2}(p_+ z + p_- \bar z)} = \label{further}\\
&& = -2 \frac{\partial}{\partial \bar z} \, 
\int \frac{d^2p}{2\pi} \hat \chi(p^2) 
e^{-\frac{1}{2}(p_+ z + p_- \bar z)} \,\,\,.
\nonumber 
\end{eqnarray}
The spectral density $\hat \chi(p^2)$ is given in terms of 
the matrix elements of the operator $\chi$ once we have factorised 
the left momentum of the intermediate states. According to the 
above discussion this is a positive quantity which depends only 
on the invariant mass square of the cluster of the intermediate 
particles. By using the identity 
\[
\hat\chi(p^2) = \int_{0}^{\infty} da^2\,
\delta(p^2 - a^2) \hat\chi(a^2)
\,\,\,, \]
eq.\,(\ref{further}) can be written then as 
\EQ
\langle \chi(z,\bar z)\chi(0,0) \rangle = 
-\frac{1}{\pi} \frac{\partial}{\partial \bar z} \,
\int_{0}^{\infty} da^2\, \hat\chi(a^2) \,K_0(a \sqrt{z\bar z})\,\,\,. 
\label{further2}
\EN 
Since 
\[
\frac{d K_0}{dx} = -K_1(x)\,\,\,,
\]
by taking the derivative of the above expression we have 
\EQ
\langle \chi(z,\bar z)\chi(0,0) \rangle = 
\frac{1}{2\pi} \sqrt{\frac{z}{\bar z}} \,
\int_{0}^{\infty} da^2 a\,\hat\chi(a^2) \,K_1(a \sqrt{z\bar z})\,\,\,. 
\label{further3}
\EN 
and therefore for the function $L$ we have 
\EQ
L(z\bar z) = \frac{(z\bar z)^{3/2}}{2\pi} \,
\int_{0}^{\infty} da^2\,a\,\hat\chi(a^2) \,K_1(a\sqrt{z\bar z})\,\,\,. 
\label{Lfinal}
\EN 
Eq.\,(\ref{Lfinal}) explicitly manifests the positivity of this function, 
Q.E.D.

The relative sum rule is easily obtained. In fact, since at the fixed 
points $H = \frac{2}{3} C$, we have 
\begin{eqnarray}
C_{in}-C_{fin} &=& \frac{3}{2} \int d(z\bar z) \,\bar z \,
\langle \chi(z,\bar z)\chi(0,0) \rangle = 
\label{fersum}\\
&& = \frac{3}{2\pi} \int_0^{\infty} da^2 \,a\,\hat\chi(a^2) 
\int_0^{\infty} dR R^2 K_1(a R) \,\,\,.\nonumber 
\end{eqnarray}
Since 
\[
\int dR \,R^2 K_1(a R) = \frac{2}{a^3} \,\,\, ,
\]
eq.\,(\ref{fersum}) can be written as 
\EQ
\Delta C = \frac{3}{\pi} \int_0^{\infty} \frac{da^2}{a^2} \hat\chi(a^2) 
\,\,\,.
\label{fersum2}
\EN 
Let us consider now the two-particle contributions of the spectral 
function $\hat\chi(a^2)$. Since 
\begin{eqnarray}
&& 
F_{bf}^{\chi}(\beta_1,\beta_2) = \frac{1}{2\cosh\frac{\beta}{2}}\,
\left(\barr\omega\, e^{-\beta_1/2}\,F_{bb}^{\Theta}(\beta) + 
\omega\, e^{-\beta_2/2} \,F_{ff}^{\Theta}(\beta) \right) \,\,\,;\\
&& F_{fb}^{\chi}(\beta_1,\beta_2) = \frac{1}{2\cosh\frac{\beta}{2}}\,
\left(\barr\omega\, e^{-\beta_2/2}\,F_{bb}^{\Theta}(\beta) - 
\omega\, e^{-\beta_1/2} \,F_{ff}^{\Theta}(\beta) \right) \,\,\,,
\nonumber 
\end{eqnarray}
we have 
\EQ
F_{bf}^{\chi} \barr F_{bf}^{\chi} + 
F_{fb}^{\chi} \barr F_{fb}^{\chi} = 
\left(e^{-\beta_1/2} + e^{-\beta_2/2}\right)\, 
\frac{\mid F_{bb}^{\Theta}\mid^2 + \mid F_{ff}^{\Theta}\mid^2}
{4 \cosh^2\frac{\beta}{2}} \,\,\,.
\label{cont}
\EN 
To obtain the relative contribution for $\hat\chi(a^2)$ we have just 
to disregard the factor 
\[
\left(e^{-\beta_1/2} + e^{-\beta_2/2}\right)
\] 
relative to the left component of the momentum, so that 
\begin{eqnarray}
\hat\chi(a^2) &=& 2\pi \int_{\beta_1 > \beta_2} 
\frac{d\beta_1}{2\pi} \frac{d\beta_2}{2\pi} \, 
\delta(a-m\cosh\beta_1-m\cosh\beta_2) \,
\delta(m\sinh\beta_1+m\sinh\beta_2)\,\times \nonumber\\
&& \,\,\,\,\,\,\,\,\,
\frac{\mid F_{bb}^{\Theta}(\beta_1-\beta_2)\mid^2 +
\mid
F_{ff}^{\Theta}(\beta_1-\beta_2)\mid^2} {4 \cosh^2\frac{\beta_1-\beta_2}{2}}
\,=\\
& =& \frac{1}{16\pi} \int_{-\infty}^{\infty} \frac{d\beta}{\cosh^3\beta} 
\delta(a-2 m \cosh\beta) \,
\left[\mid F_{bb}^{\Theta}(2\beta)\mid^2 + \mid
F_{ff}^{\Theta}(2\beta)\mid^2 \right] \nonumber 
\end{eqnarray}
Inserting this expression into eq.\,(\ref{fersum2}), we finally have 
\EQ
\Delta C^{(2)} = \frac{3}{8\pi^2 m^4} \int_0^{\infty} 
\frac{d\beta}{\cosh^4\beta} \,\left[
\mid F_{bb}^{\Theta}(2\beta)\mid^2 +
\mid F_{ff}^{\Theta}(2\beta)\mid^2 \right] \,\,\,.
\label{twofer}
\EN 
i.e. the sum rule relative to the fermionic case employs the 
same integrand as the bosonic case and therefore provides 
the same results as before.

\resection{Summary and Conclusions}

In this paper we have analysed several aspects of supersymmetric 
models from the point of view of the $S$-matrix approach. 
The simplest supersymmetric scattering theory discussed in Section 2 
has a natural interpretation and identification in terms of the 
first model described by the Roaming Series. We have determined the 
lowest Form Factors of the trace operators of the super stress-energy 
tensor and we have discussed the novelty present in this calculation. 
Several checks of their validity have been presented in Section 7, 
in particular a remarkable saturation of the $c$-theorem sum rule 
obtained for the simplest Schoutens's model. We have finally obtained 
a fermionic version of the $c$-theorem by employing the supersymmetric 
properties of the correlators. It would be extremely interesting 
to explore further the properties of these supersymmetric theories 
and obtain closed expressions for Form Factors of other operators 
and with an arbitrary number of external particles. 

\vspace{1cm}
{\em Acknowledgments}. I am extremely grateful to G. Delfino 
for useful discussions. I would like also to thank O. Babelon, 
D. Bernard, S. Penati, F. Smirnov and P. Windey for discussions 
and comments and the LPTHE for the warm hospitality during my staying, 
where part of this work has been done.

\newpage

\appendix



\appsection

In this appendix we collect some useful formulas of the function 
$G(\beta)$ used in Section 6. 

The function $G(\beta)$ is the minimal solution of the equations 
\EQ
\begin{array}{l} 
G(\beta) = U(\beta) G(-\beta) \,\,\,;\\
G(i \pi -\beta) = G(i \pi +\beta) \,\,\, ,
\end{array}
\label{unicroG1}
\EN 
where $U(\beta)$ is the function entering the $S$-matrix of the SShG 
model, eq.\,(\ref{integralU}). Its explicit expression is given by 
\EQ
G(\beta) = 
\exp\left[- \int_0^{\infty} \frac{dt}{t} 
\frac{\sinh\alpha t\,\sinh(1-\alpha)t}{\cosh^2\frac{t}{2} \sinh t \cosh t} 
\sin^2\frac{(i \pi -\beta) t}{2 \pi}\right] \,\,\,.
\label{integralG}
\EN 
For large value of $\beta$, this function tends to a constant given by 
\EQ
\lim_{\beta\rightarrow\infty} G(\beta) = 
\exp\left[- \frac{1}{2}\int_0^{\infty} \frac{dt}{t} 
\frac{\sinh\alpha t\,\sinh(1-\alpha)t}{\cosh^2\frac{t}{2} \sinh t \cosh t} 
\right] \,\,\,.
\label{constG}
\EN 
For the purposes of numerical calculation, it can be written in a 
more convenient way as follows. First of all, notice that  
the function $U(\beta)$ can be factorised as  
\EQ
U(\beta) = U_1(\beta)\, U_2(\beta) \,\,\, ,
\label{factoriseU}
\EN 
where 
\begin{eqnarray}
&& U_1(\beta) = \exp\left[-i \int_0^{\infty} \frac{dt}{t} 
\frac{(1-\cosh t)(1 + \cosh(1-2\alpha) t}{\sinh^2 t} 
\sin\frac{\beta t}{\pi}\right] \,\,\, ; \\
\label{integralU1}
&& U_2(\beta) = \exp\left[-i \int_0^{\infty} \frac{dt}{t} 
\frac{\cosh(1 - 2\alpha) t}{\cosh t} 
\sin\frac{\beta t}{\pi}\right] \,\,\, .\nonumber 
\end{eqnarray}
Correspondingly, $G$ can be factorised as 
\EQ
G(\beta) = G_1(\beta) \, G_2(\beta) \,\,\, ,
\EN 
where 
\begin{eqnarray}
&& G_1(\beta) = 
\prod_{k=0}^{N-1} 
\left[
\frac
{p^2_k(\beta,0) \,p^2_k(\beta,1-2\alpha)}
{p^2_k(\beta,1) \,p(\beta,2\alpha) \,p_k(\beta,2-2\alpha)}
\right]^{\frac{(k+1)(k+2)}{4}} \times \\
&& \,\times \exp\left[\int_0^{\infty} \frac{dt}{t} 
\frac{(1-\cosh t)(1+\cosh(1-2\alpha) t) d_N(t)}{\sinh^3 t}\,
\sin^2\frac{(i\pi-\beta) t}{2\pi} \right] \,\,\,. \nonumber 
\end{eqnarray}
and 
\begin{eqnarray}
&& G_2(\beta) = 
\prod_{k=0}^{N-1} 
\left[
\left(1 + \left(\frac{(i\pi-\beta)}{(4 k + 1 + 2 \alpha) \pi}\right)^2
\right) \,
\left(1 + \left(\frac{(i\pi-\beta)}{(4 k + 3 - 2 \alpha) \pi}\right)^2
\right) 
\right]^{\frac{1}{2}} \,\times \\
&& \,\,\,\times \,\exp\left[\int_0^{\infty} \frac{dt}{t} 
\frac{\cosh(1-2\alpha)t \,e^{-4 N t}}{\cosh t \sinh t}\,
\sin^2\frac{(i\pi-\beta) t}{2\pi} \right] \,\,\, , \nonumber 
\end{eqnarray}
with 
\[
p_k(\beta,x) = 
\left[
\left(1 + \left(\frac{(i\pi-\beta)}{(2 k + 3 + x) \pi}\right)^2
\right) \,
\left(1 + \left(\frac{(i\pi-\beta)}{(2 k + 3 - x) \pi}\right)^2
\right) 
\right] \,\,\, ,
\] 
and 
\[
d_N(t) = \frac{1}{2} 
\left[ (N+1) (N+2) - 2 N (N+2) e^{-2 t} + N (N+1) e^{-4 t}\right]\, 
e^{-2 N t} 
\] 

\appsection

In this appendix we briefly describe the reason of the inconstency of 
the Form Factors found by Ahn \cite{Ahn} for $F_{bb}^{\Theta}$ and
$F_{ff}^{\Theta}$ of the SShG model, referring the reader to his 
original paper for all notations used by this author\footnote{They 
slightly differ from ours. The reader should also be aware that 
there are several misprints in the paper \cite{Ahn}.}  

The Form Factors found by Ahn are those of formula (4.24) of 
\cite{Ahn}, i.e. 
\begin{eqnarray}
&&F_{bb}^{\Theta}(\beta) = 2\pi m^2\,\left[
\frac{(F_+^{min} + F_-^{min})}{2} + \frac{(F_+^{min} - F_-^{min})}{2} \,
\cosh\frac{\beta}{2}\right] \,\,\,;
\label{finalFFAhn} \\
&&F_{ff}^{\Theta}(\beta) = 2\pi m^2\, 
\frac{(F_+^{min} + F_-^{min})}{2} \,\sinh\frac{\beta}{2} \,\,\, ,
\nonumber
\end{eqnarray}
where 
\begin{eqnarray}
&& F_+(\beta) = \exp\left[
\int_0^{\infty} \frac{dt}{t} \frac{f_+(t)}{\sinh t} 
\,\sin^2\frac{(i \pi - \beta) t}{2\pi} \right] \,\,\,; \\
\label{FFminAhn} \\
&& F_-(\beta) = \exp\left[
\int_0^{\infty} \frac{dt}{t} \frac{f_-(t)}{\sinh t} 
\,\sin^2\frac{(i \pi - \beta) t}{2\pi} \right] \,\,\,; \nonumber 
\end{eqnarray}
and 
\EQ
f_{\pm}(t) = \frac{(1-\cosh t) (1 + \cosh((1-2 \mid \alpha\mid)t))}
{\sinh^2 t} \pm \frac{\cosh (1 - 2 \mid \alpha \mid)t)}{\cosh t} 
\label{fabsAhn}
\EN 
{\em By construction}, the functions $F_{\pm}(\beta)$ considered 
by Ahn satisfied the unitarity equations relative to the eigenvalues 
of the $S$-matrix in the $F=-1$ sector and the crossing equations 
{\em in the usual way}, namely 
\begin{eqnarray}
&& F_+(\beta + 2\pi i) = F_+(-\beta) \,\,\,; \label{crossingAhn}\\
&& F_-(\beta + 2\pi i) = F_-(-\beta) \,\,\,. \nonumber 
\end{eqnarray}
As discussed in Section 6, the crossing properties of the Form Factors 
in the $F=-1$ on the other hand are quite different. This has an
important consequence on the validity of (\ref{finalFFAhn}). Consider 
in fact the transformation $\beta \rightarrow \beta + 2\pi i$ in the 
first of (\ref{finalFFAhn}). It is easy to see that the Form Factor of the 
trace operator $\Theta$ does not satisfy in this case the equation 
\EQ
F_{bb}^{\Theta}(\beta + 2\pi i) = F_{bb}^{\Theta}(-\beta) \,\,\, ,
\EN 
expressing the locality of this operator. 

The problem becomes evident as one tries to apply the $c$-theorem sum
rule.  If this is applied by using the above formulas and equation 
(\ref{twocth}), one gets the results reported in Table 2 which 
are definitely in disagreement with the theoretical result 
$C = 3/2$. The results reported in the original Table 1 of 
the paper by Ahn were actually obtained by means of an unjustified step 
at this stage of his calculations, namely by plugging {\em negative 
values} for the quantity $\mid \alpha\mid$ which appears in all 
formulas of Ahn's paper (see, for instance eq.\,(\ref{fabsAhn}) above). 
In this case, although the first values reported by Ahn seem in 
reasonable agreement with the $c$-theorem sum rule, if that calculation 
had been pursued for higher values of the coupling constant,  
the results would have been quite unsatisfactory, as shown by the 
unreasonable small values obtained for the central charge for 
certain cases and a weird increasing behaviour around $\mid \alpha\mid
=-0.5$, see Table 3.

\newpage

\hs

\vspace{25mm}

{\bf Table Caption}

\vspace{1cm}

\begin{description}
\item [Table 1]. The two-particle contribution to the sum rule of the 
$c$-theorem for the SShG model. 
\item [Table 2]. The two-particle contribution to the $c$-theorem 
sum rule relative to the Ahn's Form Factors. 
\item [Table 3]. The two-particle contribution to the $c$-theorem 
sum rule relative to the Ahn's Form Factors for negative values 
of $\mid \alpha\mid$.  
\end{description}

\newpage

\newpage


\begin{table}

$$ \begin{array}{|c|c|c|c|}
 \hline
& &  & \\
\alpha 
\,\,\,
& \frac{\lambda^2}{4\pi} \,\,\,
& \Delta\,c^{(2)} \,\,\,
& {\makebox {\rm precision}} $\%$ \,\,\, \\
   &        &   &      \\
\hline
   &        &    &    \\
\frac{1}{100} \,\,\,& \frac{1}{99} \,\,\,& 1.49968 \,\,\,
& 0.0213 \,\,\, \\
\frac{3}{100} \,\,\,& \frac{3}{97} \,\,\,& 1.49741 \,\,\,
& 0.1726 \,\,\,\\
\frac{1}{20}\,\,\,  & \frac{1}{19}\,\,\, & 1.49349 \,\,\,
& 0.4340 \,\,\, \\
\frac{1}{10} \,\,\, & \frac{1}{9}\,\,\,  & 1.47955 \,\,\,
& 1.3633 \,\,\, \\
\frac{3}{20} \,\,\, & \frac{3}{17} \,\,\,& 1.46333 \,\,\,
& 2.4446 \,\,\, \\
\frac{1}{5} \,\,\,  & \frac{1}{4}\,\,\,  & 1.44742 \,\,\,
& 3.5053 \,\,\,\\
\frac{3}{10}\,\,\,  & \frac{3}{7} \,\,\, & 1.42109 \,\,\,
& 5.2606 \,\,\, \\
\frac{2}{5} \,\,\,  & \frac{2}{3} \,\,\, & 1.40480 \,\,\,
& 6.3466 \,\,\, \\
\frac{1}{2}\,\,\,   &  1 \,\,\,          & 1.39935 \,\,\,
 & 6.7100 \,\,\,\\
& &  &\\ \hline
\end{array}
$$
\end{table}
\begin{center}
{\bf Table 1} 
\end{center}

\newpage 

\begin{table}

$$ \begin{array}{|c|c|c|}
 \hline
& &  
 \\
\alpha 
\,\,\,
& \frac{\lambda^2}{4\pi} \,\,\,
& \Delta\,c^{(2)} \,\,\,\\
   &        &          \\
\hline
   &        &         \\
\frac{1}{100} \,\,\,& \frac{1}{99} \,\,\,& 1.51481 \,\,\,\\
\frac{3}{100} \,\,\,& \frac{3}{97} \,\,\,& 1.54328 \,\,\,\\
\frac{1}{20}\,\,\,  & \frac{1}{19}\,\,\, & 1.57019 \,\,\,\\
\frac{1}{10} \,\,\, & \frac{1}{9}\,\,\,  & 1.63080 \,\,\,\\
\frac{3}{20} \,\,\, & \frac{3}{17} \,\,\,& 1.68227 \,\,\,\\
\frac{1}{5} \,\,\,  & \frac{1}{4}\,\,\,  & 1.72524 \,\,\,\\
\frac{3}{10}\,\,\,  & \frac{3}{7} \,\,\, & 1.78833 \,\,\,\\
\frac{2}{5} \,\,\,  & \frac{2}{3} \,\,\, & 1.82442 \,\,\,\\
\frac{1}{2}\,\,\,   &  1 \,\,\,          & 1.83616 \,\,\,\\
& &  \\ \hline
\end{array}
$$
\end{table}
\begin{center}
{\bf Table 2} 
\end{center}

\newpage 

\newpage 

\begin{table}

$$ \begin{array}{|c|c|}
 \hline
& 
 \\
\mid\alpha\mid 
\,\,\,
& \Delta\,c^{(2)} \,\,\,\\
           &          \\
\hline
           &         \\
-0.001 \,\,\,& 1.4985 \,\,\,\\
-0.005 \,\,\,& 1.49246 \,\,\,\\
-0.01 \,\,\, & 1.48485 \,\,\,\\
-0.02 \,\,\, & 1.46951 \,\,\,\\
-0.03 \,\,\, & 1.45413 \,\,\,\\
-0.05 \,\,\, & 1.42369 \,\,\,\\
-0.1 \,\,\,  & 1.35190  \,\,\,\\
-0.15\,\,\,  & 1.28762 \,\,\,\\
-0.20  \,\,\,& 1.23038 \,\,\,\\
-0.33  \,\,\,& 1.14079 \,\,\,\\
-0.40. \,\,\,& 1.14552 \,\,\,\\
-0.50. \,\,\,& 1.58825 \,\,\,\\
 &  \\ \hline
\end{array}
$$
\end{table}
\begin{center}
{\bf Table 3} 
\end{center}

\newpage

\hs 

\vspace{25mm}

{\bf Figure Captions}

\vspace{1cm}

\begin{description}
\item [Figure 1]. Renormalization Group Flows described by the Roaming 
Models. 
\item [Figure 2]. Bound states and residues in some amplitudes of the
Schoutens's $S$-matrix. 
\item [Figure 3]. Complex zeros of the Roaming
Models. 
\item [Figure 4]. Logarithm of the correlation function $m^{-4} \langle
\Theta(x) \Theta(0) \rangle$ versus $m r$ for $\alpha=0$ (full line) and for 
$\alpha=0.5$ (dashed line). 
\item [Figure 5]. Bound state recursive equation for the form factor 
${\cal F}_2$. 
\item [Figure 6]. Correlation function $m^{-4} \langle
\Theta(x) \Theta(0) \rangle$ versus $m r$ of the first Schoutens's model,
obtained  with the first two terms of the spectral series.  
\end{description}

\newpage

\begin{figure}
\centerline{
\psfig{figure=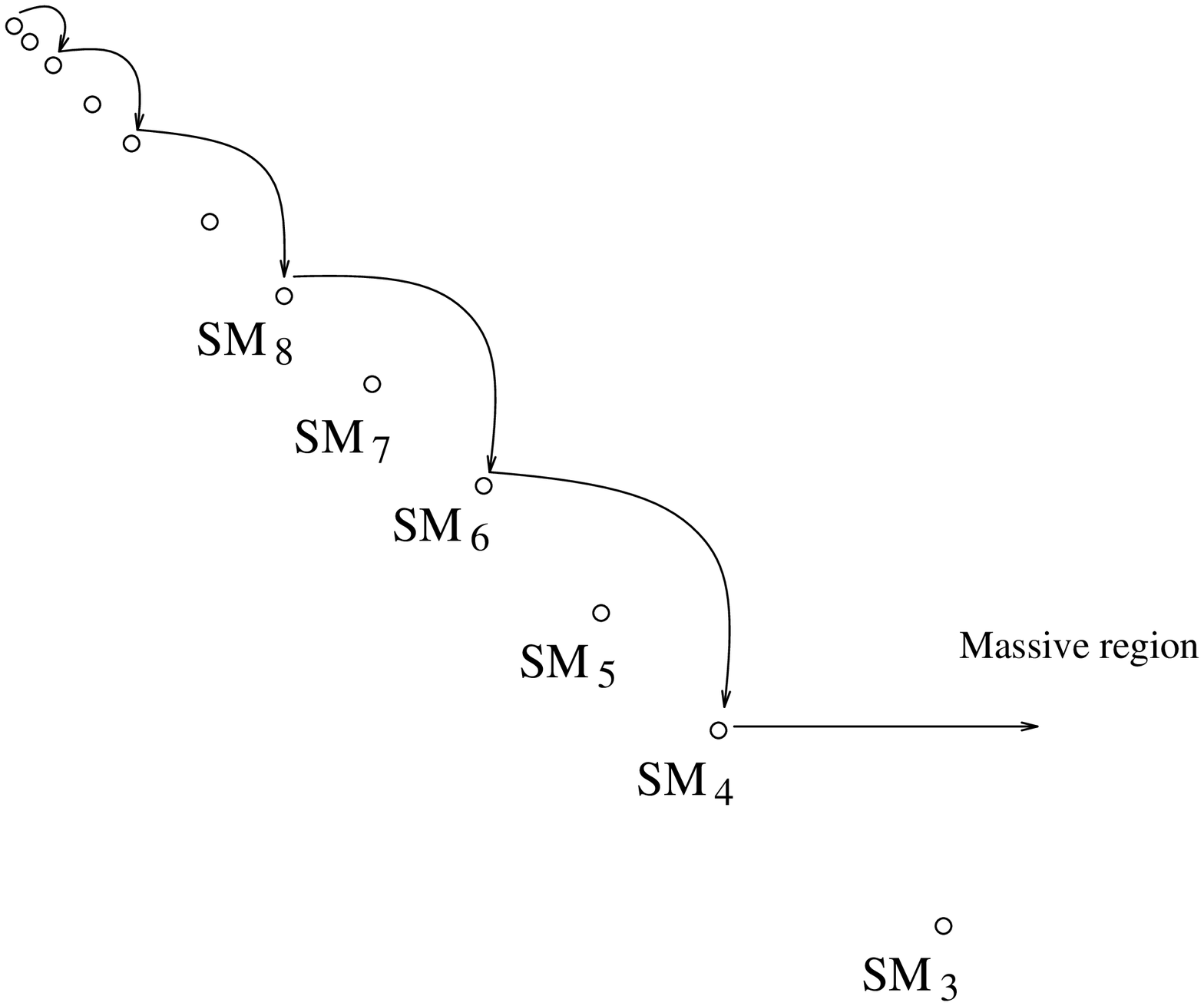}}
\vspace{-1cm}
\begin{center}
{\bf Figure 1}
\end{center}  
\end{figure}

\newpage

\begin{figure}
\centerline{
\psfig{figure=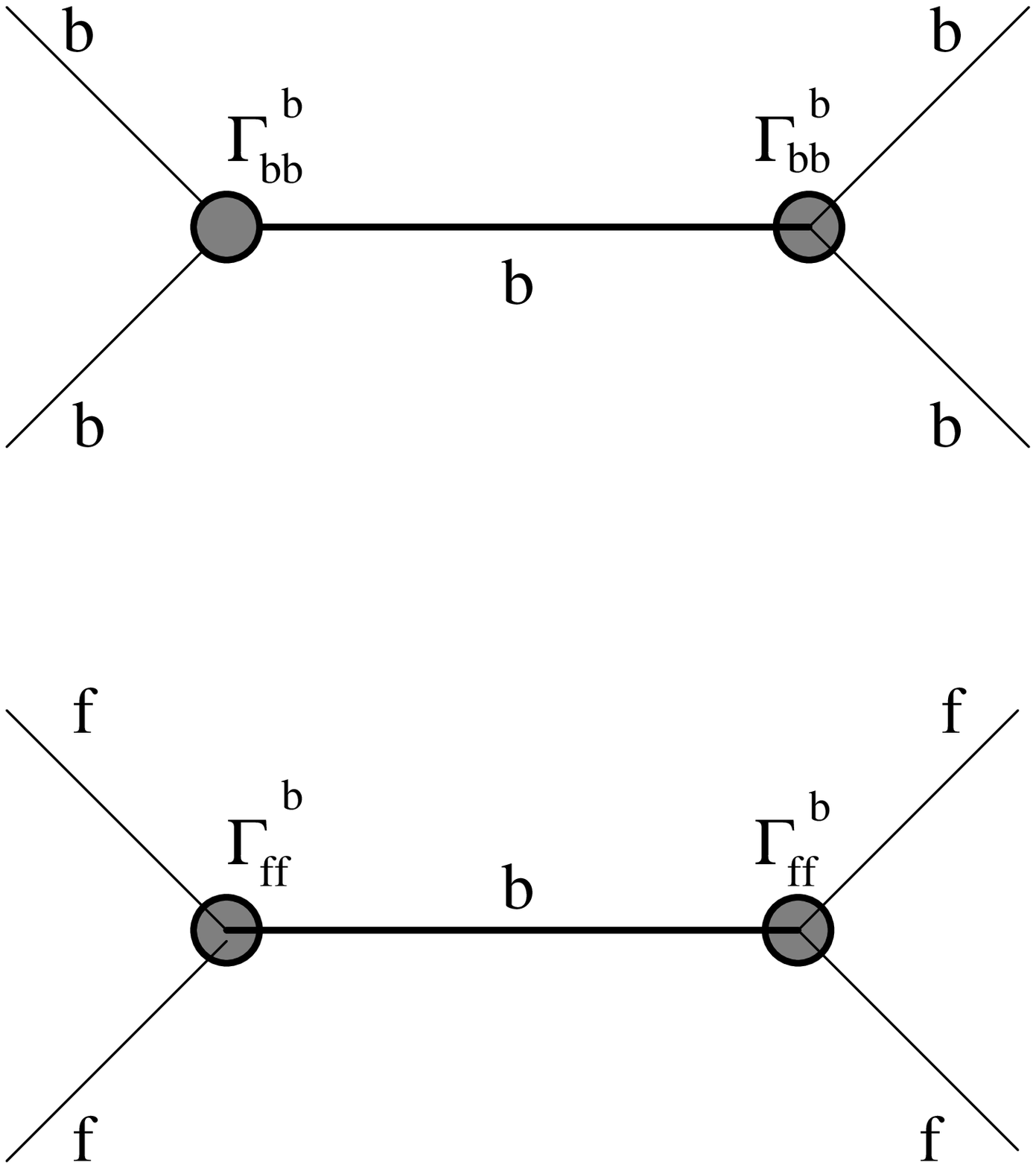}}
\vspace{-1cm}
\begin{center}
{\bf Figure 2}
\end{center}  
\end{figure}  

\newpage 

\begin{figure}
\centerline{
\psfig{figure=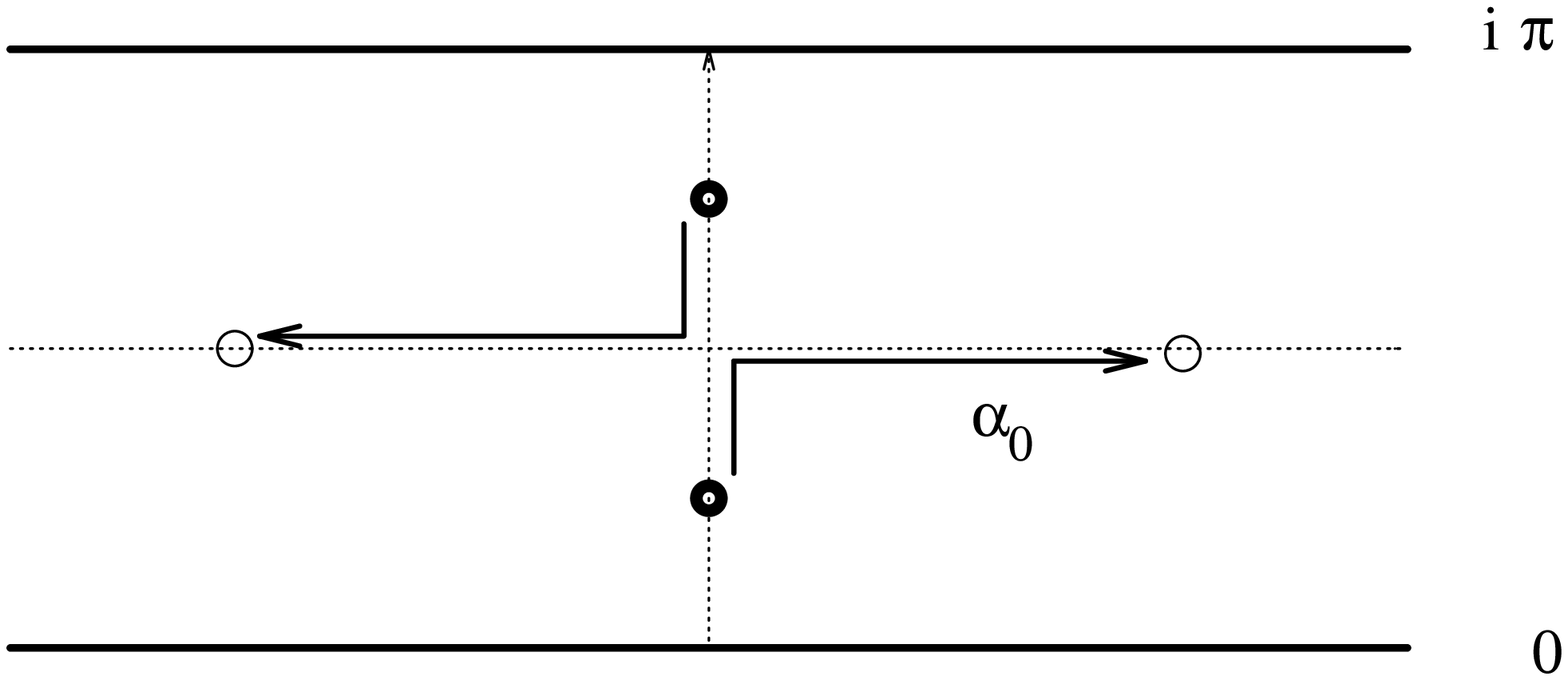}}
\vspace{1cm}
\begin{center}
{\bf Figure 3}
\end{center}  
\end{figure}  

\newpage 

\begin{figure}
\centerline{
\psfig{figure=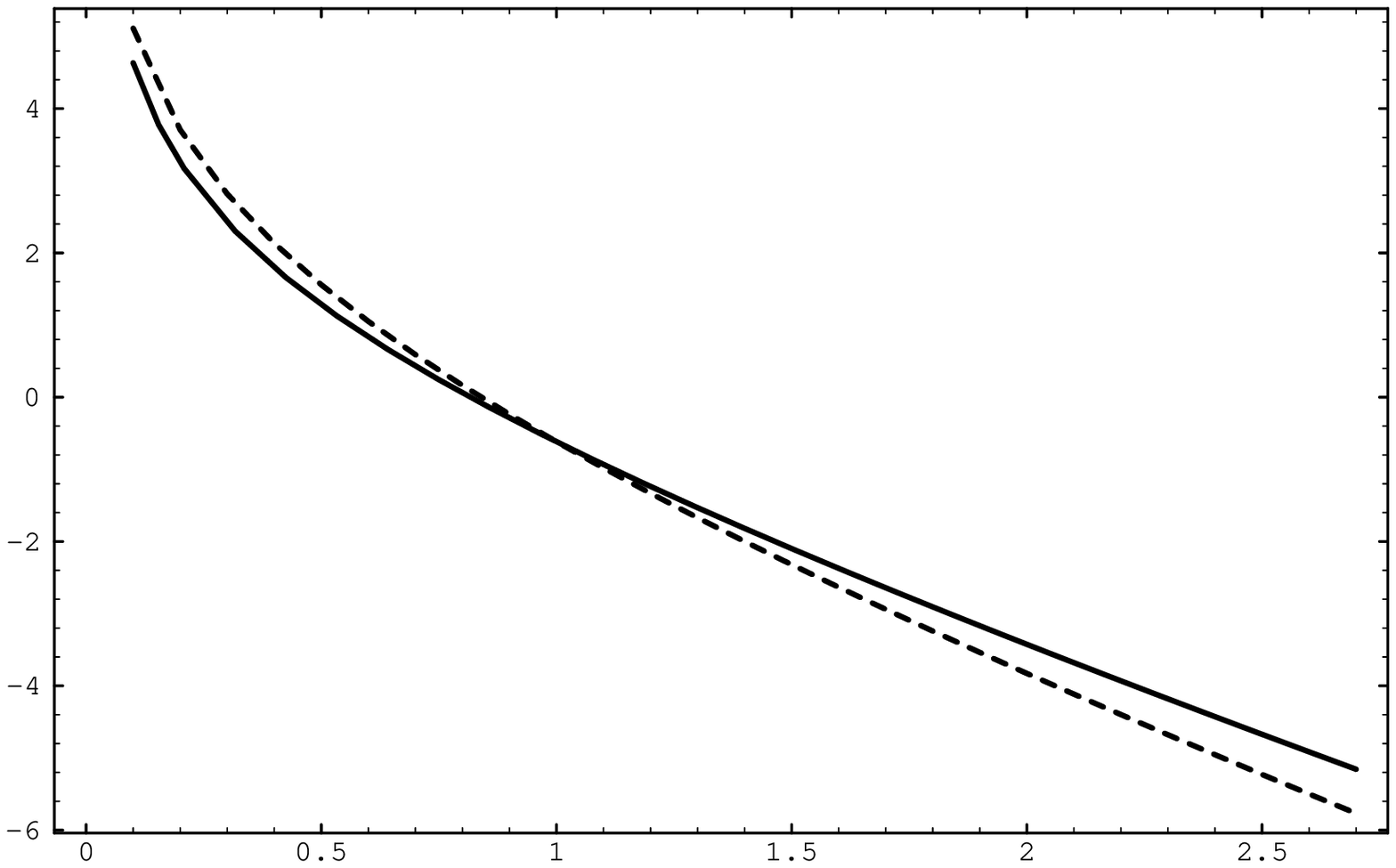}}
\vspace{-1cm}
\begin{center}
{\bf Figure 4}
\end{center}  
\end{figure}  

\newpage 

\begin{figure}
\centerline{
\psfig{figure=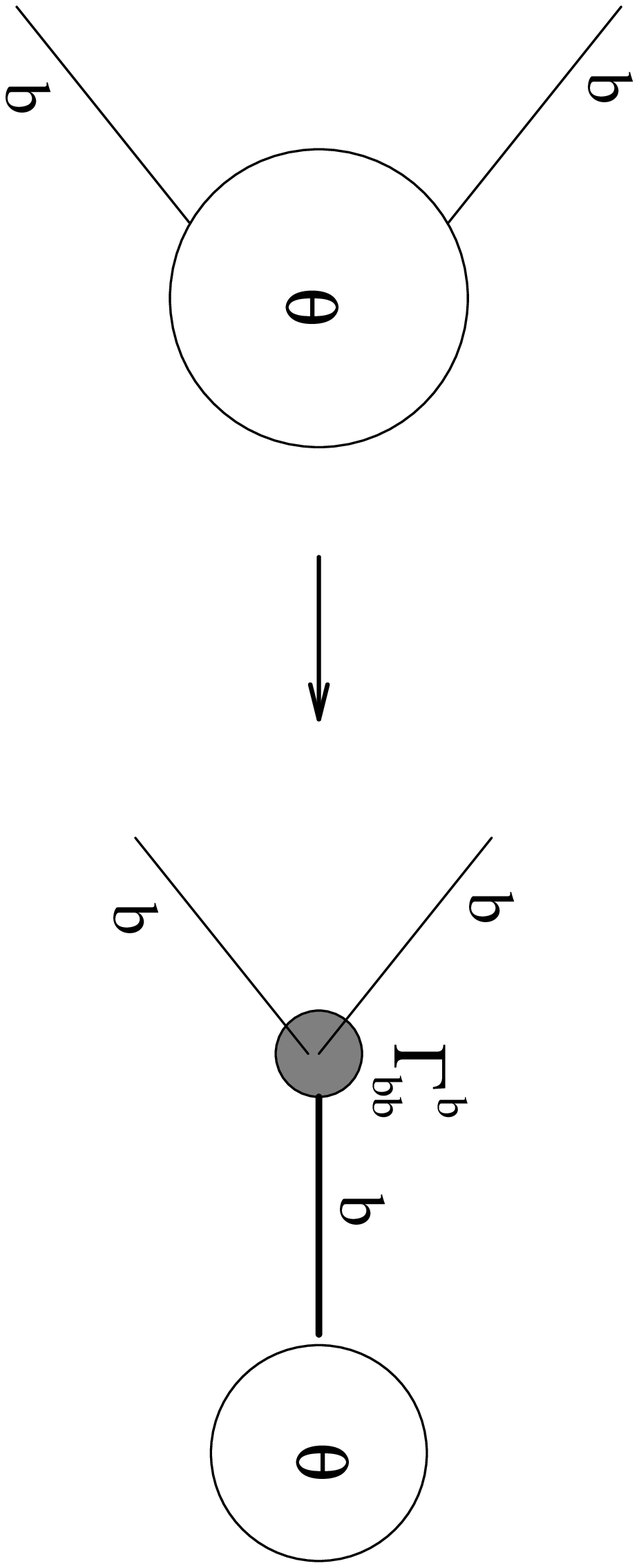}}
\vspace{-1cm}
\begin{center}
{\bf Figure 5}
\end{center}  
\end{figure}  

\newpage 

\begin{figure}
\centerline{
\psfig{figure=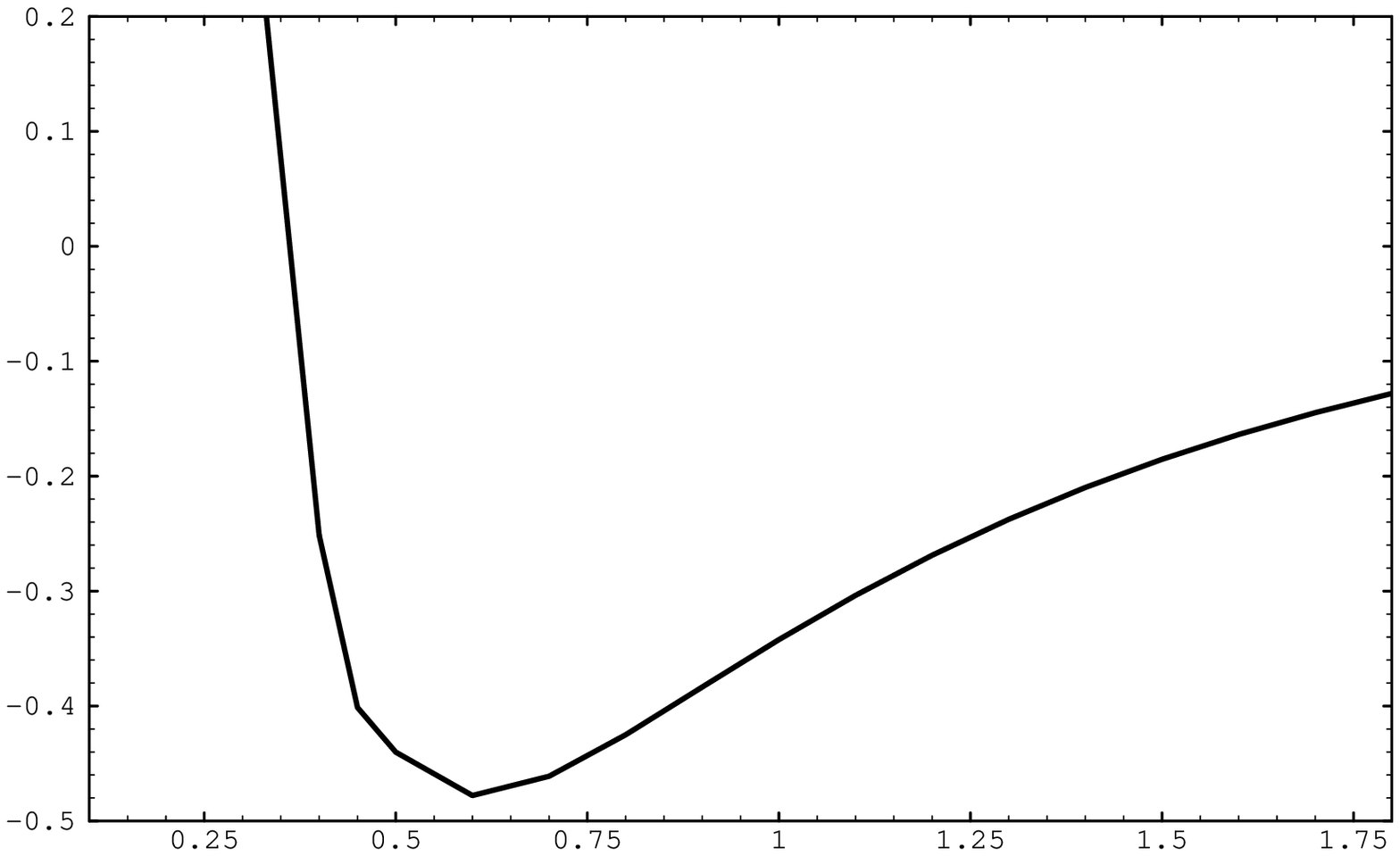}}
\vspace{-1cm}
\begin{center}
{\bf Figure 6}
\end{center}  
\end{figure}  




\end{document}